\documentclass[reprint,preprintnumbers,nofootinbib,superscriptaddress,amsmath,amssymb,aps,prd]{revtex4-2}
\pdfoutput=1

\usepackage{booktabs}
\usepackage{framed}
\usepackage{graphicx}
\graphicspath{{figures/}}
\usepackage{hyperref}
\usepackage{siunitx}
\usepackage{xcolor}

\hypersetup{colorlinks=true, linkcolor=blue, citecolor=blue}

\renewcommand{\vec}[1]{\mathbf{#1}}
\newcommand{\unitvec}[1]{\hat{\mathbf{#1}}}
\newcommand{\ev}[1]{\left\langle#1\right\rangle}
\newcommand{\abs}[1]{\left\lvert {#1} \right\rvert}
\newcommand{\md}{\mathrm{d}}
\newcommand{\me}{\mathrm{e}}
\newcommand{\Sb}{\vec{S}_{\text{bkg}}}

\begin{document}

\title{Charting the Skyrmion Free-Energy Landscape}
\preprint{IPPP/23/11}
\preprint{MITP-23-012}

\author{Juan Carlos Criado}
\email{jccriadoalamo@ugr.es}
\affiliation{CAFPE and Departamento de Física Teórica y del Cosmos, Universidad de Granada, Campus de Fuentenueva, E–18071 Granada, Spain}

\author{Peter D.~Hatton}
\email{p.d.hatton@durham.ac.uk}
\affiliation{Centre for Materials Physics, Department of Physics, Durham University, South Road, Durham DH1 3LE, United Kingdom}

\author{\'Alvaro Lanza}
\email{alanzaserrano@gmail.com}
\affiliation{Department of Physics, Durham University, Durham DH1 3LE, United Kingdom}

\author{Sebastian Schenk}
\email{sebastian.schenk@uni-mainz.de}
\affiliation{PRISMA$^{+}$ Cluster of Excellence \& Mainz Institute for Theoretical Physics, Johannes Gutenberg-Universit\"at
Mainz, 55099 Mainz, Germany}

\author{Michael Spannowsky}
\email{michael.spannowsky@durham.ac.uk}
\affiliation{Institute for Particle Physics Phenomenology, Department of Physics, Durham University, South Road, Durham DH1 3LE, United Kingdom}

\begin{abstract}
Chiral magnets with Dzyaloshinskii–Moriya interactions feature a rich phase diagram with a variety of thermodynamical phases.
These include helical and conical spin arrangements and topologically charged objects such as (anti)skyrmions.
Crucially, due to hysteresis effects, the thermodynamical phases can co-exist at any given temperature and external magnetic field, typically leading to metastability of, e.g., the material's topological phase.
In this work, we use Monte Carlo simulations to study these effects.
We compute the relative free energies of co-existing states, enabling us to determine the ground state at all values of the external parameters.
We also introduce a method to estimate the activation energy, i.e.~the height of the energy barrier that separates the topological phase from the ground state.
This is one of the key ingredients for the determination of the skyrmion lifetime, which is relevant for technological applications.
Finally, we prescribe predicting the system's evolution through any path in the space of external parameters.
This can serve as a guideline to prepare the magnetic material in any desired phase or even trigger a phase transition in an experimental setup.
\end{abstract}

\maketitle

\section{Introduction}
\label{sec:intro}

Chiral magnets are magnetic materials typically dominated by ferromagnetic exchange and Dzyaloshinskii–Moriya (DM)~\cite{Dzyaloshinskii:1958,Moriya:1960zz} interactions.
They exhibit a topological phase, in which a triangular lattice of partially topologically-protected structures, known as skyrmions, is stabilised~\cite{Skyrme:1962vh,Bogdanov:1989,Bogdanov:2020}.
This phase has been observed experimentally in MnSi~\cite{Muhlbauer:2009} for the first time, and in several other compounds since then~\cite{Munzer:2009var, yu2010real, yu2011near, tokunaga2015new, woo2016observation, fujima2017thermodynamically}.
Apart from the general interest in exploring the fundamental properties of skyrmions, they are studied for their potential technological applications in magnetic data storage technology~\cite{Fert:2013}, racetrack memory~\cite{tomasello2014strategy}, artificial synapses for neuromorphic computing~\cite{song2020skyrmion}, reservoir computing~\cite{pinna2020reservoir}, and reshuffling for signal decorrelation in probabilistic computing~\cite{zazvorka2019thermal}.
The counterpart of skyrmions with opposite topological charge, known as antiskyrmions, have also been studied experimentally~\cite{koshibae2016theory, hoffmann2017antiskyrmions, huang2017stabilization, camosi2018micromagnetics, kovalev2018skyrmions, bottcher2018b, jena2020elliptical} and have their own set of applications in spintronics~\cite{huang2017stabilization}.

In practice, the stabilisation of skyrmions strongly depends on the system's external parameters, such as temperature or applied magnetic field (see, e.g.,~\cite{cortes2017thermal,buttner2018theory,birch2021topological}).
As a function of these parameters, chiral magnets feature a rich phase diagram with various thermodynamical phases, including helical and conical spin arrangements as well as the topological phase.
Typically, the transition between these phases is subject to hysteresis effects, which can lead to the co-existence of different phases at the same temperature and external magnetic field, depending on the previous evolution of the system.
Crucially, as one would expect the system to feature a unique ground state, this implies that, for instance, the topological phase can be metastable.
In this case, the skyrmions would have a finite lifetime.
It is challenging to determine the true ground state of the system by distinguishing it from the co-existing phase, in particular in scenarios close to a phase transition (see, e.g.,~\cite{PhysRevLett.96.047207, PhysRevLett.107.037207, PhysRevB.89.094411}).
We address this problem in this work.

Monte Carlo algorithms provide a powerful tool in the theoretical study of thermodynamic phases of magnetic materials~\cite{do2009skyrmions, yu2010real, Buhrandt:2013uma, Criado:2021gzp, Araz:2022cdm}.
In particular, it has been shown that these techniques can reproduce a thermodynamical phase diagram as measured in experiments~\cite{Buhrandt:2013uma}.

\begin{figure}
  \centering
  \includegraphics[width=0.75\columnwidth]{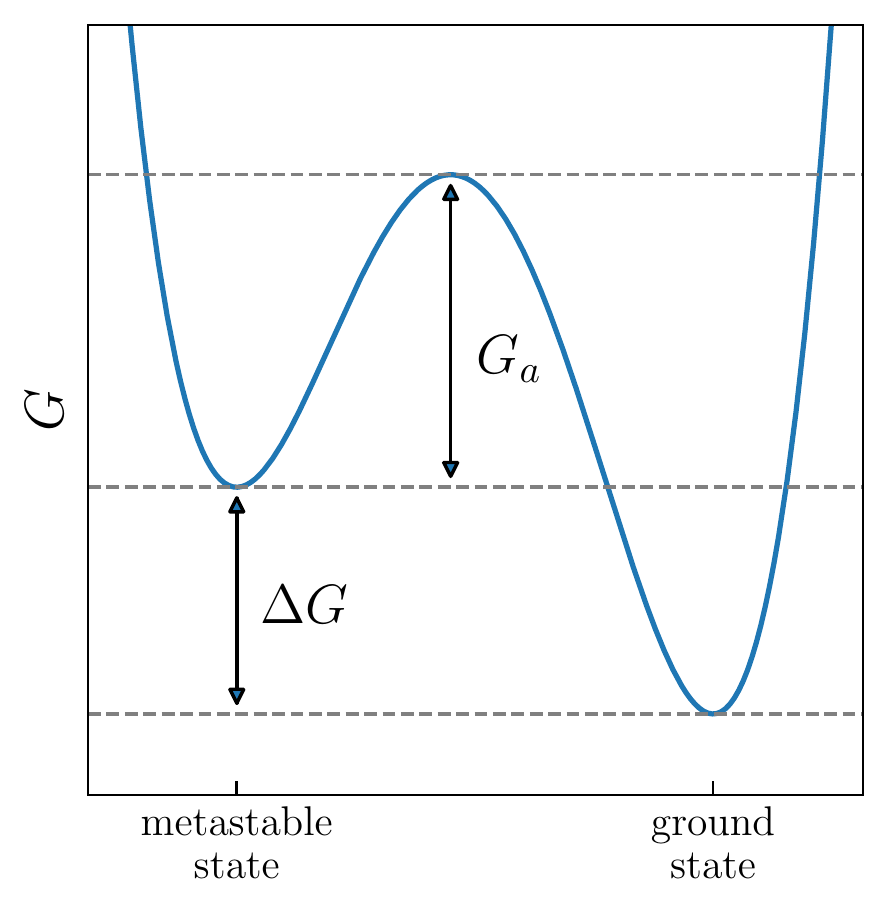}
  \caption{Schematic illustration of the Gibbs free energy $G$ as the system is forced to transition from a metastable state to the ground state. The necessary activation energy $G_a$ corresponds to the height of the barrier separating both. The difference in free energies between both vacua is denoted by $\Delta G$.}
  \label{fig:energy-schematic}
\end{figure}

As we will point out in this work, this even includes the hysteresis effects mentioned before.
In this work, we investigate the structure of the phase diagram for chiral magnets that support either Bloch skyrmions\footnote{A Bloch skyrmion is a vortex in which the projection of the spins into the plane perpendicular to the external magnetic field is orthogonal to the radial direction.} or antiskyrmions.
In particular, we focus on points where several thermodynamical phases co-exist and study their properties.
These properties include the spectrum of Helmholtz free energies, $F$, associated with each phase.
By means of free energy, we can determine the system's true ground state at each point in the phase diagram.
We also compute the Gibbs free energy $G$ as a function of the magnetisation.
In particular, we obtain the difference $\Delta G$ between the topological phase and the $xy$-translation invariant phase and the activation energy $G_a$, corresponding to the height of the Gibbs free energy barrier between them.
For every point in the system's parameter space, the ground state is one of these two phases.
A schematic illustration of the profile of $G$ and the two quantities $G_a$ and $\Delta G$ is shown in Fig.~\ref{fig:energy-schematic}.
Both are crucial for understanding the metastability of the topological skyrmion phase, which has been investigated in, e.g.,~\cite{Munzer:2009var, 10.1126/science.1234657, BESSARAB2015335, 10.1038/nmat4752, 10.1038/ncomms12669, 10.1038/nphys3506, 10.1126/sciadv.1701704, 10.1038/s41467-017-01353-2, 10.1126/sciadv.1602562, 10.1103/PhysRevMaterials.1.074405, 10.1021/acs.nanolett.6b04821, PhysRevB.98.134407, birch2019increased, PhysRevB.102.104416, PhysRevB.105.214435}.
In particular, $G_a$ represents the minimal energy that must be provided to the system to transition from the topological phase to the ground state if the topological phase is metastable.
By the Eyring equation, it is intimately related to the skyrmion lifetime~\cite{bessarab2018lifetime, PhysRevB.101.060403, muckel2021magnetic, lobanov2021stability, voronin2022activation},
\begin{equation}
  \tau \propto \frac{1}{\beta} \me^{\beta G_a} \, .
  \label{eq:eyring}
\end{equation}
Here, $\beta = \left(k_B T\right)^{-1}$ denotes the inverse temperature.
That is, at fixed temperature, the lifetime grows exponentially with the activation energy.
Most of the studies in this context have been done for thin 2D slices.
In this work, we consider a 3D bulk material.

To explore the spectrum of free energies associated with the thermodynamical phases, we introduce a method that allows tracking and comparing the free energy values as the magnetic field is varied in Monte Carlo simulations.
Our method is based on numerical thermodynamic integration to obtain the free energy profile for a given phase and use the detailed balance equation to assemble free energy profiles from different phases.
We also address how to predict the system's state after it is prepared in a given initial state and the temperature and magnetic field are changed quasi-statically following any specific path.
We study this question by observing the phase transitions along several paths and obtaining a general rule for their prediction.

The rest of this work is organised as follows.
In Section~\ref{sec:simulations}, we briefly describe the interactions of chiral magnets and their Monte Carlo treatment as a discrete spin-lattice system.
We then determine the thermodynamical phase diagram in Section~\ref{sec:hierarchy}.
In particular, we obtain the spectrum of free energies associated with the different phases for DM interactions that stabilise Bloch skyrmions and antiskyrmions.
In Section~\ref{sec:gibbs}, we demonstrate how the activation energy can be obtained to study the topological phase's metastability.
Furthermore, Section~\ref{sec:predicting} presents a general prescription for predicting the system's evolution as the temperature and magnetic field are changed.
Finally, we briefly summarise our results and conclude in Section~\ref{sec:conclusions}.

\section{Monte Carlo simulations of chiral magnets}
\label{sec:simulations}

\begin{table*}
  \centering
  \begin{tabular}{cccc}
    \toprule
    Point group
    & $\operatorname{DM}(\vec{M})$
    & $\operatorname{DM}_d(\vec{S})_{\vec{r}}$
    & Topological structure
    \\
    \midrule
    $T$ or $O$
    & $\mathbf{M} \cdot (\nabla \times \mathbf{M})$
    & $\mathbf{S_r} \cdot \left(
      \mathbf{S}_{\mathbf{r} + \hat{\mathbf{x}}} \times \hat{\mathbf{x}}
      + \mathbf{S}_{\mathbf{r} + \hat{\mathbf{y}}} \times \hat{\mathbf{y}}
      + \mathbf{S}_{\mathbf{r} + \hat{\mathbf{z}}} \times \hat{\mathbf{z}}
      \right)
      $
    & Bloch skyrmion
    \\
    \midrule
    $D_{2d}$
    & $\mathbf{M} \cdot (\partial_x \mathbf{M} \times \hat{\mathbf{x}} - \partial_y \mathbf{M} \times \hat{\mathbf{y}})$
    &
      $
      \vec{S}_{\vec{r}} \cdot (\vec{S}_{\vec{r} + \unitvec{y}} \times \unitvec{x}
      -  \vec{S}_{\vec{r} + \unitvec{x}} \times \unitvec{y})
      $
    & Antiskyrmion
    \\
    \bottomrule
  \end{tabular}
  \caption{Possible DM interactions allowed by the crystal structure (i.e.~point group) of the material. Here, $\operatorname{DM}(\vec{M})$ illustrates the interaction in terms of the continuous magnetisation field, while $\operatorname{DM}_d(\vec{S})_{\vec{r}}$ denotes its discretised counterpart in a spin-lattice system.}
  \label{tab:DM}
\end{table*}

In characterising bulk chiral magnets, we employ a coarse-grained description in which a continuous vector field $\vec{M}$ describes the material's magnetisation.
Below the Curie temperature, the norm of this vector is fixed and equal to the saturation magnetisation, $\abs{\vec{M}} = M_s$.
In this framework, the associated Hamiltonian of the chiral magnet is then given by
\begin{equation}
 H
 =
 \int \md^3r \,
 \left[
   \frac{J}{2} \left(\nabla \vec{M}\right)^2
  + K \operatorname{DM}(\vec{M})
  - \vec{B} \cdot \vec{M}
 \right] \, ,
 \label{eq:hamiltonian}
\end{equation}
where $J$ and $K$ are free parameters, and $\operatorname{DM}(\vec{M})$ denotes the DM interaction.
The latter is generally odd under space inversions and linear in both $\vec{M}$ and $\partial_i \vec{M}$.
However, their specific form varies across different materials.
In this work, we consider DM interactions that arise in materials with $T$, $O$ and $D_{2d}$ crystal structure, shown in Table~\ref{tab:DM}.

For a Monte Carlo treatment of this system, we discretize space in a cubic lattice of size $30 \times 30 \times 30$ with neighbouring points separated by a distance $a$.%
\footnote{%
  The simulation of this 3D system is necessary to recover some of the observed features of chiral magnets~\cite{Buhrandt:2013uma}, but it incurs a significant computational cost in comparison with the more typical simulations of 2D lattices. We address this by parallelizing our simulations in a GPU device, achieving a factor $\sim 10^4$ speed up, as described in Ref.~\cite{Criado:2021gzp}.
}
We further impose periodic boundary conditions.
In terms of the unit vector field $\vec{S} = \vec{M} / M_s$, the lattice Hamiltonian reads
\begin{equation}
\begin{split}
 H_d
 =
 - \sum_{\vec{r}} \Big[
 &
  \tilde{J} \; \mathbf{S}_{\vec{r}} \cdot \left(
   \mathbf{S}_{\vec{r} + \unitvec{x}}
   + \mathbf{S}_{\vec{r} + \unitvec{y}}
   + \mathbf{S}_{\vec{r} + \unitvec{z}}
  \right)
 \\
 &
 	+ \tilde{K} \operatorname{DM}_d \left( \vec{S} \right)_{\vec{r}}
  + \tilde{B} \left(\vec{S}_{\vec{r}}\right)_z
 \Big] \, .
\end{split}
\label{eq:discrete-hamiltonian}
\end{equation}
Here, we assume the external magnetic field is parallel to the positive $z$-direction, $\vec{B} = B_z \hat{z}$.
Furthermore, we have defined the couplings $\tilde{J} = J M_s^2 a$, $\tilde{K} = K M_s^2 a^2$, and $\tilde{B} = B_z M_s a^3$, all having dimensions of energy.
The discretised DM interactions $\operatorname{DM}_d(\vec{S})_{\vec{r}}$ are shown in Table~\ref{tab:DM}.

At finite temperature $T$, the probability distribution of the system's microstates is characterised by the Boltzmann distribution.
The thermal expectation value of any operator $\mathcal{O}$, in the presence of a source $\vec{J}$, is then given by a Boltzmann-weighted integral over all states,
\begin{equation}
  \ev{\mathcal{O}}_{\vec{J}} = \frac{1}{Z} \int \mathcal{D}\vec{S} \,
  \mathcal{O} \me^{- \left(H_d + \vec{J} \cdot \vec{S} \right) / \tilde{T}} \, ,
  \label{eq:ev}
\end{equation}
where we have defined $\tilde{T} = k_B T$.
By construction, the source is a vector field, and
$
\vec{J} \cdot \vec{S} = \sum_{\vec{r}} \vec{J}_{\vec{r}} \cdot \vec{S}_{\vec{r}}.
$
Similarly, the generating functional $Z$, and also the Helmholtz free energy $F$ of the system, are defined by
\begin{equation}
  Z[\mathbf{J}] = e^{-F[\mathbf{J}] / \tilde{T}} = \int \mathcal{D}\vec{S} \,
  \me^{-\left(H_d + \mathbf{J} \cdot \mathbf{S}\right) / \tilde{T}} \, .
  \label{eq:Z}
\end{equation}
Before we continue, we remark that all quantities, $\ev{\mathcal{O}}$, $Z$, and $F$, are invariant under a rescaling of the parameters of the form
\begin{equation}
  \tilde{J} \to \lambda \tilde{J} \, , \enspace
  \tilde{K} \to \lambda \tilde{K} \, , \enspace
  \tilde{B} \to \lambda \tilde{B} \, , \enspace
  \tilde{T} \to \lambda \tilde{T} \, ,
\end{equation}
for some arbitrary constant $\lambda$, and under a simultaneous rescaling of the (arbitrary) source.
Without loss of generality, one can hence fix the value of $\tilde{J}$ to unity, $\tilde{J} = 1$.
For the rest of this work, we provide all values of the parameters $\tilde{K}$, $\tilde{B}$ and $\tilde{T}$ in these units.\footnote{Alternatively, one may view this choice as replacing these parameters by their dimensionless versions, $\tilde{K} / \tilde{J}$, $\tilde{B} / \tilde{J}$, and $\tilde{T} / \tilde{J}$.}

In the following, to obtain the thermodynamical phase diagram of the chiral magnet, we are interested in the thermal expectation values of the normalised magnetisation field at vanishing source, $\ev{\vec{S}} \equiv \ev{\vec{S}}_{\vec{J}= 0}$.
We obtain these through Monte Carlo sampling, as outlined below.

\subsection{Obtaining thermal expectation values}
\label{sec:thermal}

To determine the thermal expectation value of the normalised magnetisation field, $\ev{\vec{S}}$, we implement a Metropolis-Hastings algorithm~\cite{Metropolis:1953am, Hastings:1970aa}.
In each step of the algorithm, a random node of the spin lattice is selected, and a new spin is proposed with a uniform probability distribution over the set of unit vectors.
The selected spin is then updated with probability
\begin{equation}
  p\left(\Delta H_d\right) = \min \left\{1, e^{-\Delta H_d / \tilde{T}}\right\},
  \label{eq:acceptance}
\end{equation}
where $\Delta H_d$ is the change in energy induced by the spin update.

A lattice \emph{sweep} is defined by performing as many single-spin updates as the number of lattice nodes.
In our case, one sweep amounts to $27,000$ spin updates.
To simulate an adiabatic process in which the temperature and magnetic field are changed slowly, we update them in small increments, performing $\num{e4}$ thermalisation sweeps at each value.
Then, after a suitable thermalisation, we obtain the expectation value of an observable by computing its average over $2000$ configurations.
The latter is obtained by performing $50$ sweeps between one configuration and the next.

When moving through the configuration space of temperature and magnetic field, we refer to the path followed by $\tilde{T}$ and $\tilde{B}$ during a simulation as a \emph{schedule}.
We consider the following basic schedules designed to match the experimental procedures:
\begin{enumerate}
  \item \textit{Zero-field cooling (ZFC).} Starting at $\tilde{B} = 0$ and $\tilde{T} = 2$, decrease $\tilde{T}$ to its target value in 20 steps, following a geometric progression. Then, increase $\tilde{B}$ to its target value in 20 equally-spaced steps.
  \item \textit{High-field cooling (HFC).} Starting at $\tilde{B} = 0.6$ and $\tilde{T} = 2$, decrease $\tilde{T}$ to its target value in 20 steps, following a geometric progression. Then, decrease $\tilde{B}$ its target value in 20 equally-spaced steps.
  \item \textit{Field cooling (FC).} Starting at $\tilde{B}$ fixed at its target value and $\tilde{T} = 2$, decrease $\tilde{T}$ to its target value in 20 steps, following a geometric progression.
\end{enumerate}
The finite size of the lattice induces anisotropy effects that can be partially corrected by introducing next-to-nearest neighbour interactions, as described in Appendix~\ref{sec:nnn-corrections} and in Refs.~\cite{Martinelli:1982db,Niedermayer:1996eb}.
We implement these corrections in all our simulations.
To speed them up, we divide the lattice into three sublattices forming a three-dimensional checkerboard-like pattern, with the property that the spins in a given sublattice do not interact.
The calculation can then be accelerated by simultaneously updating all the spins in a sublattice (see, e.g.,~\cite{romero2019performance}).
This reduces the simulation time by a factor of about $\mathcal{O} \left(\num{e4}\right)$, compared to a serial implementation of the Metropolis-Hastings algorithm.

In all our simulations, we fix $\tilde{K} = \tan(2\pi/10)$.
It can be shown that this leads to helical configurations with a period of 10 lattice sites~\cite{do2009skyrmions}.
This period is directly related to the skyrmion radius~\cite{Criado:2021gzp}, and in this case it allows for the lattice we use to contain up to 9 skyrmions, when they are densely packed.

It has been shown that the results obtained with this procedure closely mimic the experimental scenario, including metastable states and hysteresis~\cite{Criado:2021gzp}.
The latter effectively amounts to a dependence of the thermal expectation value on the schedule, i.e.~the path that the Markov chain Monte Carlo samples in configuration space.
Let us briefly elaborate on this subtlety.

\subsection{Path dependence of thermal expectation values}

In an ideal scenario, the thermal expectation values of operators, as well as the free energy and partition function of the system, are entirely determined by the external parameters $\tilde{J}$, $\tilde{K}$, $\tilde{B}$ and $\tilde{T}$.
However, they also depend on how the system is prepared in practice.
The reason is that the Hamiltonian $H_d$ is generally a non-convex function of $\vec{S}$.
Depending on the path followed by the system in the external-parameter space, it might become trapped in the proximity of a local minimum of $H_d$, with a temperature $\tilde{T}$ lower than the energy barrier that separates the local minimum from the global one.

This path dependence leads to hysteresis effects, in which the system can stay in a metastable state (close to a local minimum) for long periods.
This has been confirmed experimentally (see, e.g.,~\cite{Munzer:2009var}).
When this happens, the integration space in Eqs.~\eqref{eq:ev} and~\eqref{eq:Z} reduces from the space of all states to only those close to the selected local minimum.
Therefore, to capture this phenomenon, the path integration should be done over the spin configurations that belong to the same macrostate $A$.
Equivalently, one can keep the entire integration space and change the integration measure instead, i.e.~schematically one can write
\begin{equation}
  \mathcal{D}\vec{S} \to \mathcal{D}\vec{S} \; \vec{1}_A(\vec{S}),
  \label{eq:partial-measure}
\end{equation}
where $\vec{1}_A$ is the indicator function with support in $A$,
\begin{equation}
	\vec{1}_A \left(\vec{S}\right) = 
	\begin{cases}
		1 & \text{if } \vec{S} \in A \\
		0 & \text{otherwise}
	\end{cases} \, .
\end{equation}

The quantities $\ev{\mathcal{O}}$, $Z$ and $F$ then become multi-valued functions of the external parameters, with each branch computed by integration with the measure in Eq.~\eqref{eq:partial-measure}.
These are then labeled by the macrostate $A$, denoted as $\ev{\mathcal{O}}^{(A)}$, $Z^{(A)}$ and $F^{(A)}$.
In adiabatic processes, a particular branch of these functions is selected by continuity over the path followed by the system in the external-parameter space.
This will become important in our following analysis of the phase diagram of a chiral magnet.

\section{Spectrum of metastable states}
\label{sec:hierarchy}

To study the phase diagram for the two types of DM interactions displayed in Table~\ref{tab:DM}, we simulate the system's evolution through each of the schedules defined in Section~\ref{sec:thermal}, with the target temperature and magnetic field given by the desired values of the temperature and magnetic field.
We then compute the expectation value $\ev{\vec{S}_{\vec{r}}}$ of the spin configuration by averaging over 2000 configurations separated by 50 sweeps.
We find several distinct phases of the system across these simulations.
They can be identified through the structure presented by $\ev{\vec{S}_{\vec{r}}}$ as follows.
\begin{enumerate}
  \item \textit{The topological phase}, characterized by the presence of tube-like structures, invariant under translations along the $z$ axis, and having topological charge $|Q| \simeq 1$, with
  \begin{equation}
    Q = \frac{1}{4 \pi} \sum_{\vec{r}} \ev{\vec{S}_{\vec{r}}} \cdot
    \left(\ev{\vec{S}_{\vec{r} + \unitvec{x}}} \times \ev{\vec{S}_{\vec{r} + \unitvec{y}}}\right),
    \label{eq:charge}
  \end{equation}
  where the sum is over all lattice nodes in the $xy$ plane, for any fixed value of $z$. Structures with $Q \simeq -1$ are known as skyrmions, whereas those with $Q \simeq 1$ are antiskyrmions.
They are supported by materials both with $T$ or $O$ point group and $D_{2d}$ point group, respectively.
        We find that, apart from the phase boundaries, the (anti)skyrmions are densely packed, forming a hexagonal structure.
        The lattice size and value of $\tilde{K}$ we have selected allow for an hexagonal arrangement of 9 skyrmion tubes in this phase.

\begin{figure*}
  \centering
  \includegraphics[width=0.5\columnwidth]{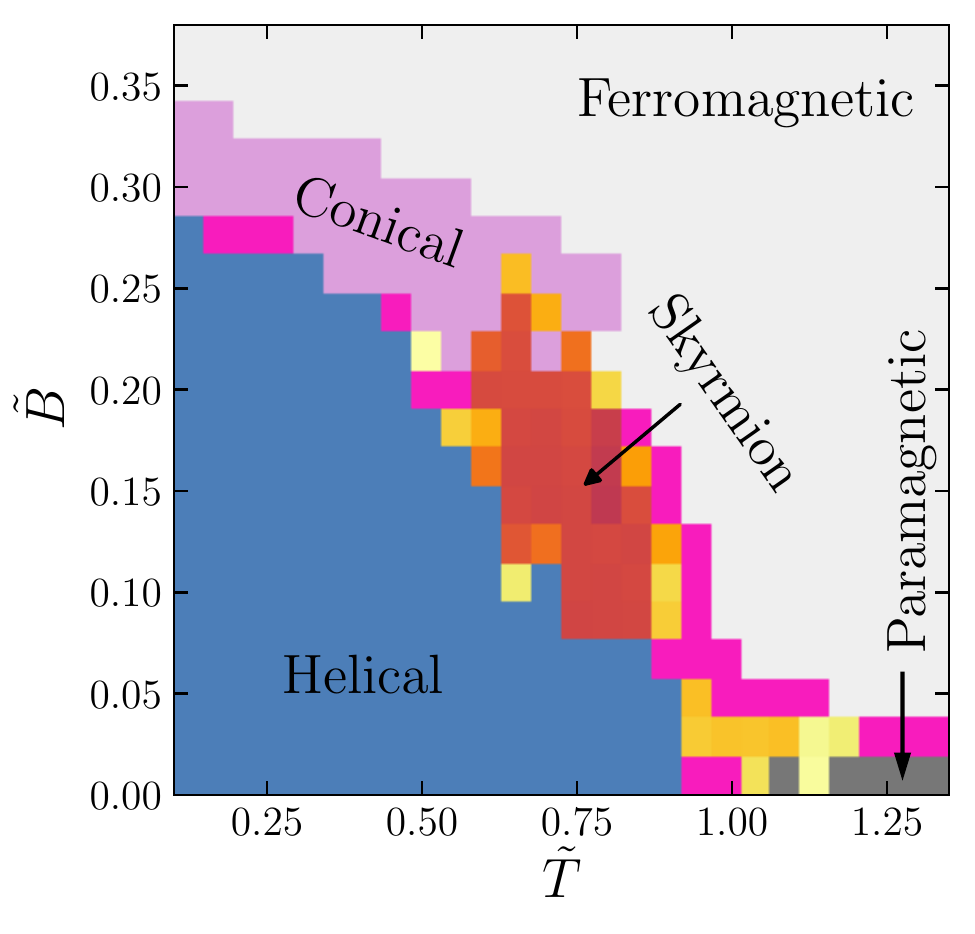}
  \includegraphics[width=0.5\columnwidth]{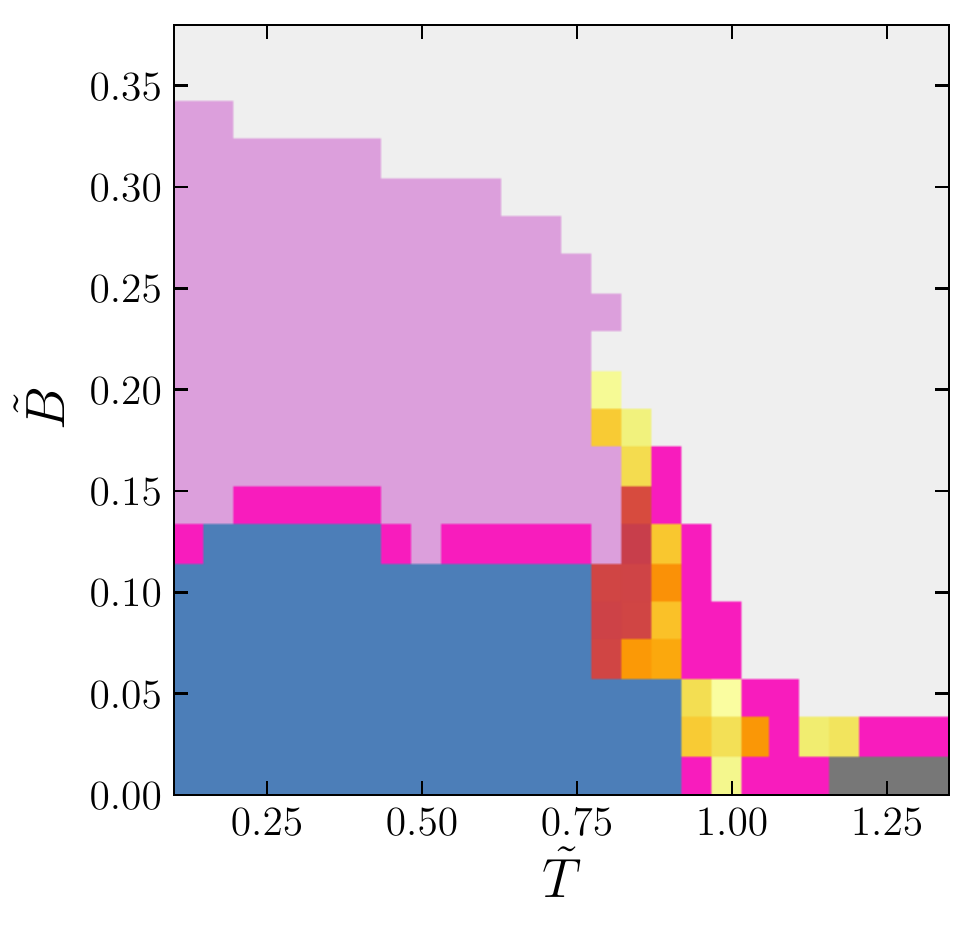}
  \includegraphics[width=0.5\columnwidth]{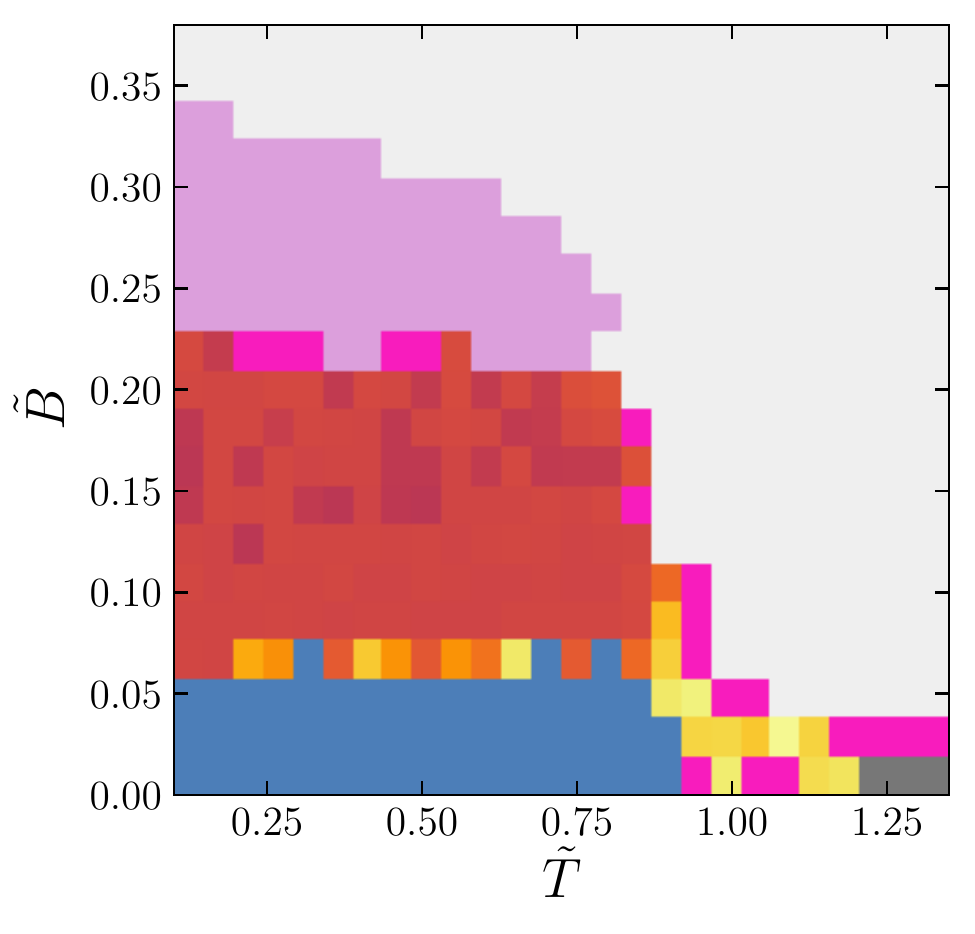}
  \includegraphics[width=0.5\columnwidth]{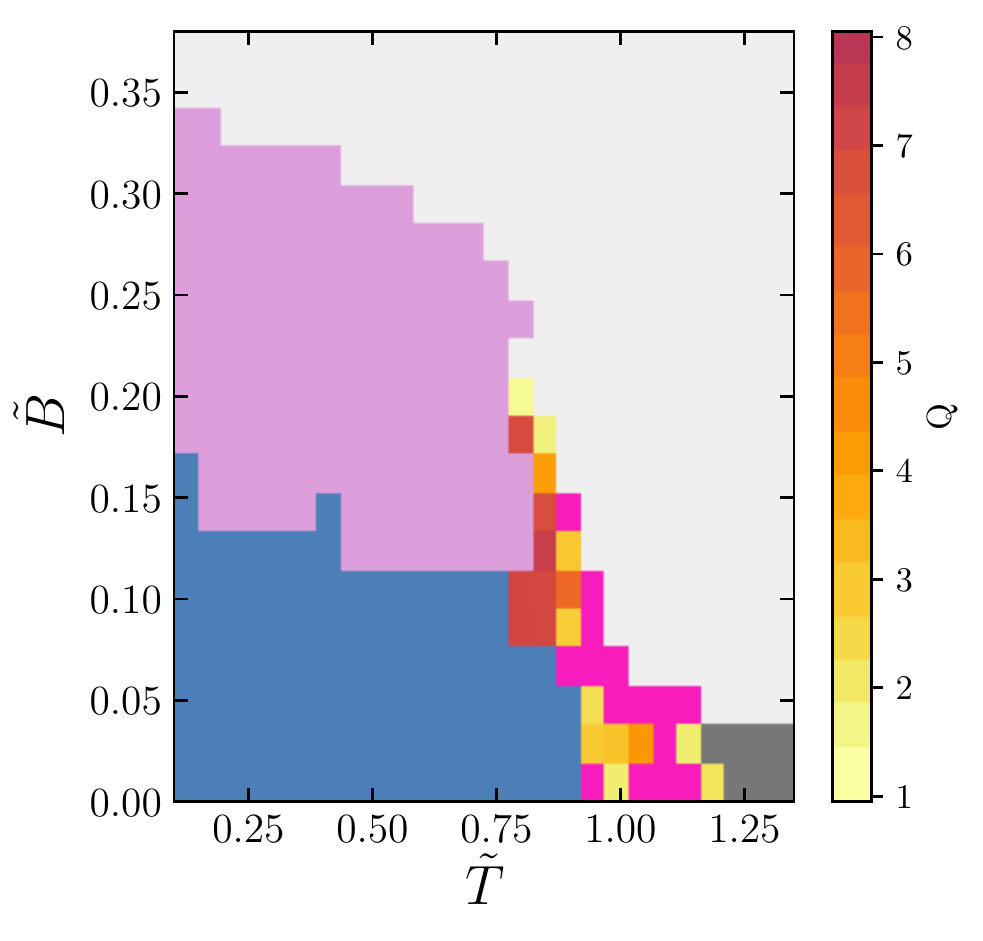}
  \includegraphics[width=0.5\columnwidth]{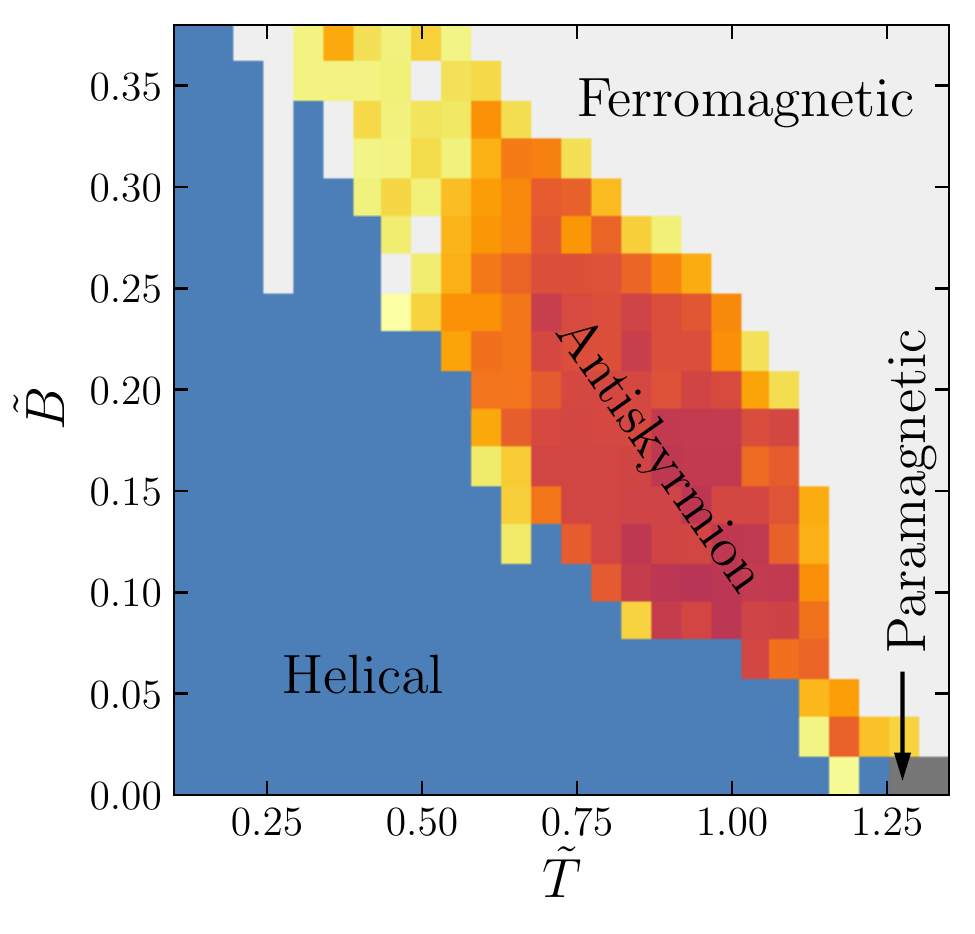}
  \includegraphics[width=0.5\columnwidth]{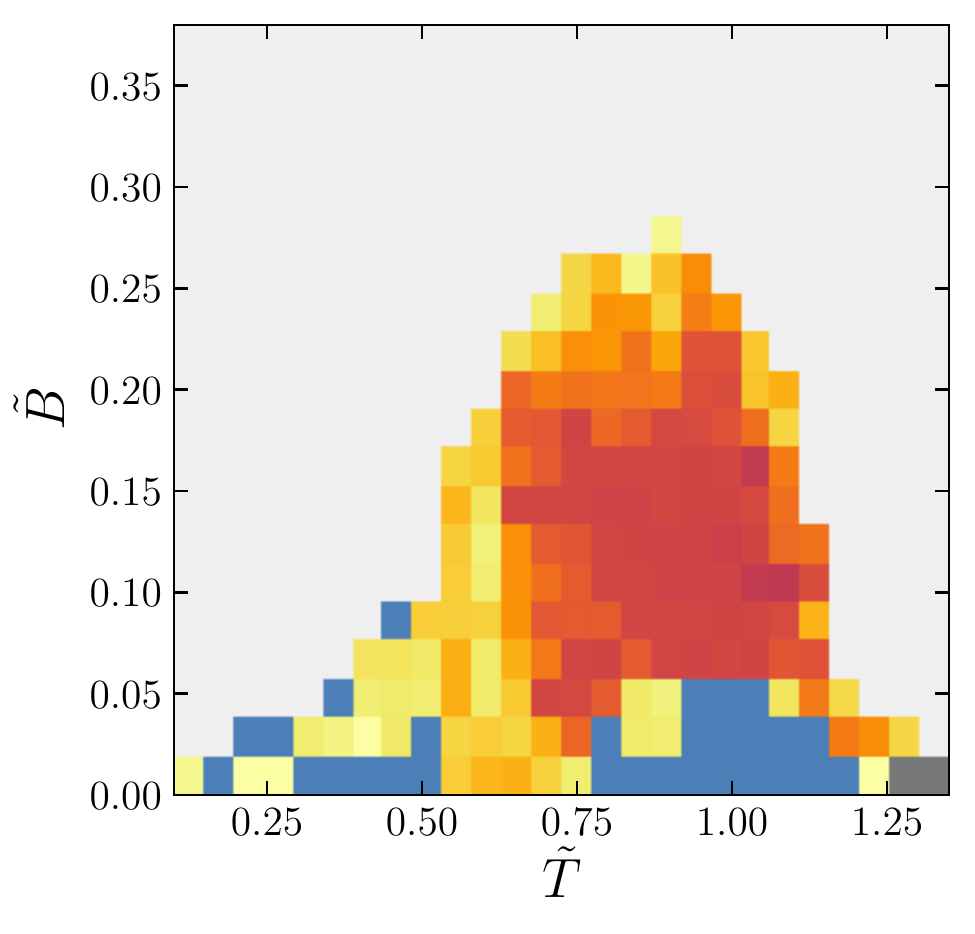}
  \includegraphics[width=0.5\columnwidth]{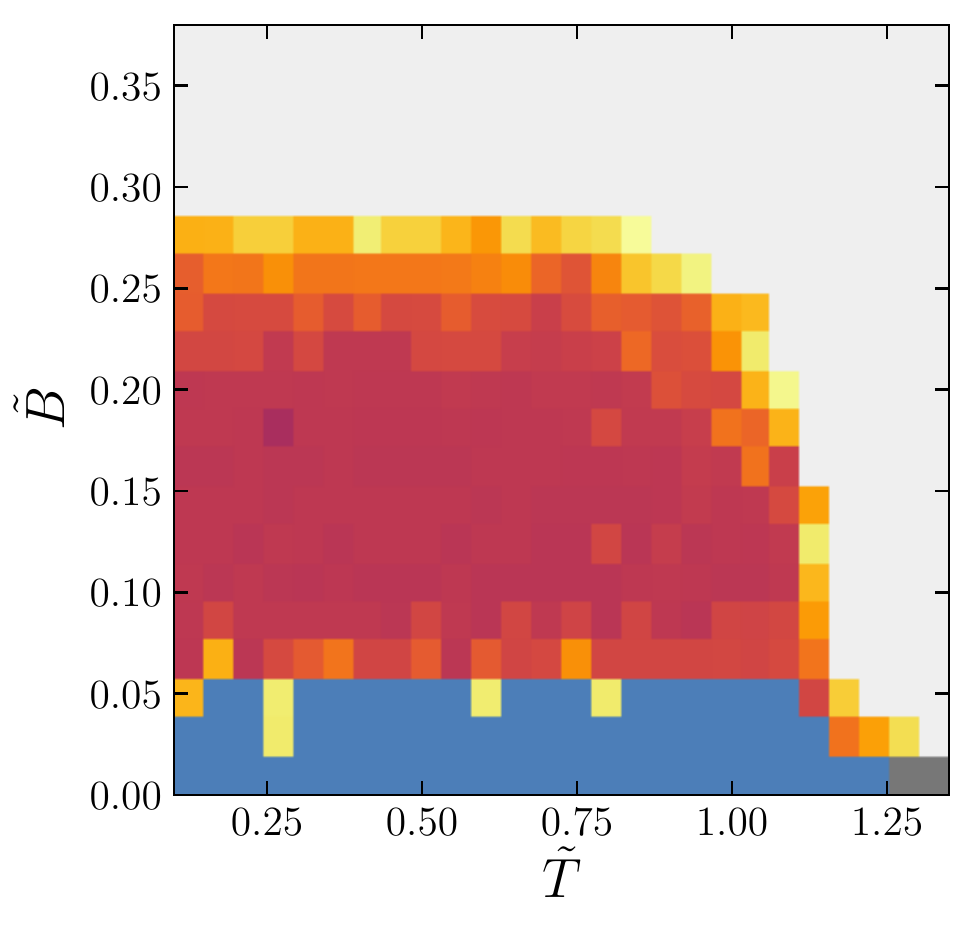}
  \includegraphics[width=0.5\columnwidth]{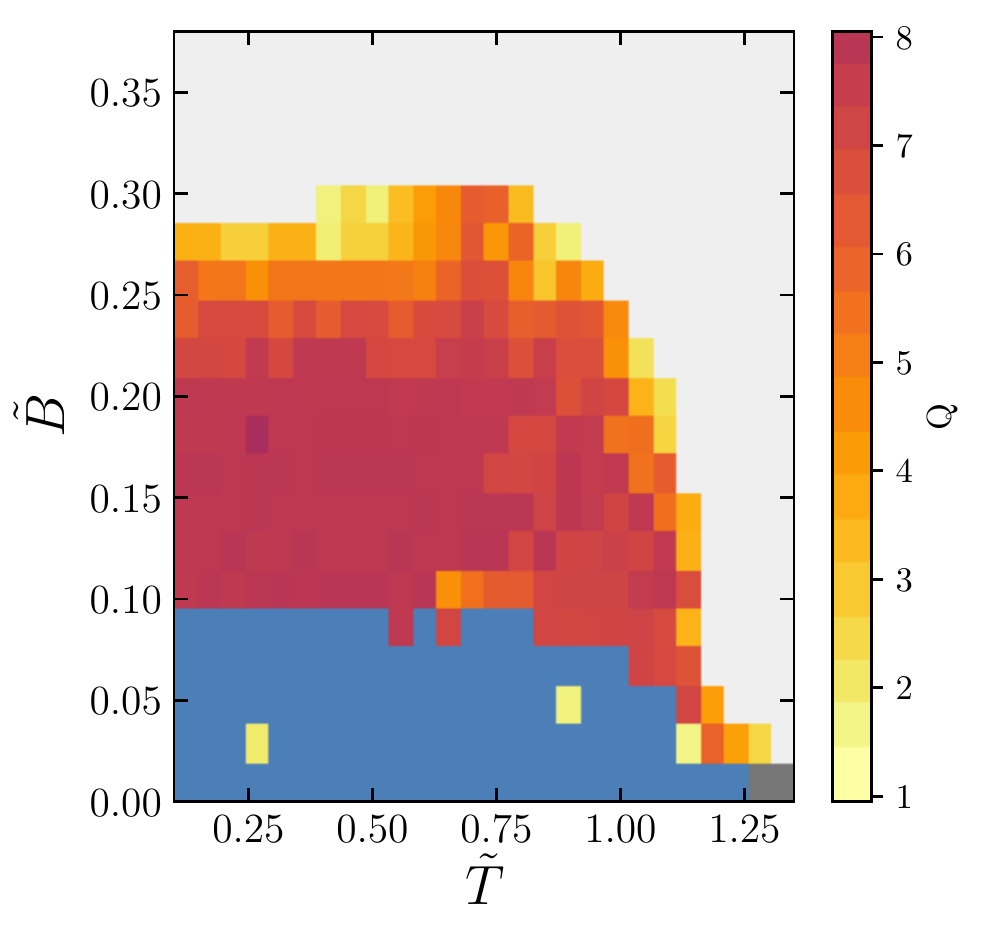}
  \caption{Thermodynamical phase diagrams for materials with $T$ or $O$ (top), featuring Bloch skyrmions, and $D_{2d}$ (bottom) crystal symmetry, featuring antiskyrmions, as a function of temperature $\tilde{T}$ and magnetic field $\tilde{B}$. The first three panels illustrate the ZFC, HFC, and FC schedules (from left to right) described in the main text. The rightmost panels show the corresponding ground state of the system. In all panels the topological (anti)skyrmion phase is shown in yellow-red, where the colorbar illustrates the total topological charge $Q$, from low (yellow) to large charge (red). Furthermore, the helical phase is shown in blue, the conical phase is shown in purple, and the ferromagnetic (paramagnetic) phase is shown in light (dark) grey. In the first three columns, points where multiple phases co-exist are shown in magenta. For a description of these, see the main text. In the rightmost panels, the magenta color coding denotes states of non-trivial structures featuring a topological charge, which, however, are distinct from ordered skyrmion tubes, lacking a $z$-translation invariance.}
  \label{fig:phase-diagram}
\end{figure*}

  \item \textit{The helical phase}, with an easy axis in the $\ev{111}$ directions for the $T$ or $O$ materials, and in any direction in the $xy$ plane for the $D_{2d}$ materials. The spins are orthogonal to the easy axis and rotate uniformly as one moves along it. There is translational symmetry along any direction perpendicular to the easy axis.

  \item \textit{The conical phase,} which only appears in $T$ or $O$ materials. There is symmetry under translations along the $xy$ plane. There is a constant $z$ component for all spins. The spins rotate uniformly around the $z$ axis as one moves along it.

  \item \textit{The ferromagnetic phase}, where all spins are aligned with the magnetic field.

  \item \textit{The paramagnetic phase}, in which there is no particular order in the spins.
\end{enumerate}
We can identify the different phases by the methods outlined in Ref.~\cite{Buhrandt:2013uma}.
Similar methods have been used to compute $\Delta F_{ab}$ in 2D materials in Ref.~\cite{PhysRevB.103.214417}.
Firstly, we classify the configuration as being in the topological phase if (and only if) $\abs{Q} \gtrsim 1$.
If $\abs{Q} < 1$, we compute the Fourier transform of the spin expectation value,
\begin{equation}
  \label{eq:1}
  \ev{\vec{S}_{\vec{k}}} = \frac{1}{N} \sum_{\vec{r}} e^{i\vec{k} \cdot \vec{r}} \ev{\vec{S}_{\vec{r}}} \, ,
\end{equation}
with $N$ being the number of lattice sites.
The phase can then be determined through the number and position of separate peaks in the intensity spectrum, $\abs{\ev{\vec{S}_{\vec{k}}}}^2$.
In the helical phase, two Bragg peaks corresponding to the wavevector direction are \textit{diagonally} present in all Fourier planes (but only the $k_{xy}$ in the $D_{2d}$ models due to the lack of $z$ axis DMI modulation for this symmetry).
In the conical phase, the peaks are \textit{vertically} placed in the $k_{xz}$ and $k_{yz}$ planes, with no contribution in $k_{xy}$.
Ferromagnetic configurations only produce one peak at the centre.

In practice, we encounter classification difficulties near the conical-helical boundary in ZFC and HFC phase diagrams, where one phase continuously distorts into the other along the temperature axis.
For instance, in a given $T, O$ symmetry configuration in this regime, we observe Bragg peaks in the $k_{yz}$ plane only (implying a conical configuration), yet with a horizontal alignment (suggesting helical).
In this case, we manually inspect the phase, marking such points of ambiguity on the phase diagram in magenta.

We show each phase diagram for the ZFC, HFC, and FC schedules in Fig.~\ref{fig:phase-diagram}.\footnote{The $D_{2d}$ scenario has already been presented in~\cite{Criado:2021gzp}.}
We find a topological phase pocket for the three schedules and both types of DM interactions.
Thus, the topological phase is the only available thermal state in the region where the pockets for the three schedules intersect.
At temperatures below the ones in the intersection region, different phases are generated by different schedules for the same final temperature and magnetic field, $\tilde{T}$ and $\tilde{B}$, respectively.
These are, therefore, the metastable states described in Section~\ref{sec:simulations}.
At the boundaries between phases, we observe mixed states of two or more of these.

To identify the true ground state of the system, we need to determine the free-energy difference between co-existing phases.
That is, we obtain
\begin{equation}
	\Delta F_{ab} = F^{(b)}\rvert_{\vec{J} = 0} - F^{(a)}\rvert_{\vec{J} = 0} \, ,
\end{equation}
between the zero-source free energies of two (meta)stable states $a$ and $b$ that arise at any temperature and magnetic field by following the method outlined in Ref.~\cite{Buhrandt:2013uma}.
One can perform simulations of the system in both states to compute the average transition probabilities between $p(a \to b)$ and $p(b \to a)$.
These determine
\begin{equation}
  e^{-\Delta F_{ab} / \tilde{T}} = \frac{p(b \to a)}{p(a \to b)} \, .
  \label{eq:acc-ratio}
\end{equation}
In practice, we find that $\tilde{T}$ is much smaller than the energy difference of both states, $\ev{H}_{\vec{J} = 0}^{(b)} - \ev{H}_{\vec{J} = 0}^{(a)}$.
This leads to the approximate relation
\begin{equation}
  \Delta F_{ab} \simeq \ev{H}_{\vec{J} = 0}^{(b)} - \ev{H}_{\vec{J} = 0}^{(a)},
  \label{eq:energy-diff-simple}
\end{equation}
Across all the phase diagrams for the $D_{2d}$ crystal structure, the different phases at a given point are ordered with the following hierarchy,
\begin{align}
  F_{\rm helical} < F_{\rm topological} < F_{\rm FM} & \quad \text{for } \tilde{B} \lesssim 0.1 \, , \\
  F_{\rm topological} < F_{\rm helical} < F_{\rm FM} & \quad \text{for } 0.1 \lesssim \tilde{B} \lesssim 0.3 \, , \\
  F_{\rm FM} < F_{\rm topological} & \quad \text{for } 0.3 \lesssim \tilde{B} \, ,
\end{align}
in the region where these phases co-exist, and where $F_p$ denotes the free energy for the phase $p$ (and FM denotes the ferromagnetic phase).

On the other hand, for the $T$ or $O$ crystals supporting Bloch skyrmions, we find
\begin{equation}
  F_{\rm conical} < F_{\rm helical} < F_{\rm topological} \, ,
  \label{eq:sk-inequalities}
\end{equation}
in the  $0.14 \lesssim \tilde{B} \lesssim 0.20$ range at low temperatures, the only region where all three phases co-exist. Otherwise, the ordering changes throughout the parameter space, and is easily deduced from the ground state diagram in Fig.~\ref{fig:phase-diagram}.
Unlike the $D_{2d}$ case, neither conical or topological phases persist at higher fields above the upper boundary, with the ferromagnetic phase being the only stable option.

These relations allow us to determine the state with minimal free energy for each temperature and magnetic field.%
\footnote{%
  While dedicated methods to find the ground state efficiently, such as parallel-tempering Monte Carlo~\cite{bottcher2018b}, already exist, we have designed our method so that we can also keep the meta-stable states, in which we are interested, and compute their relative free-energies and study the transitions between them.
}
We refer to this state as the ground state.
In Fig.~\ref{fig:phase-diagram}, we include a diagram showing the ground state at every point.
We remark that for DM interactions of type $T$ or $O$, this diagram closely resembles the one obtained in Ref.~\cite{Buhrandt:2013uma}, especially pertaining to the small temperature range spanned by the topological phase pocket.

\section{The Gibbs free energy profile}
\label{sec:gibbs}

\begin{figure*}
  \centering
  \includegraphics[width=0.65\columnwidth]{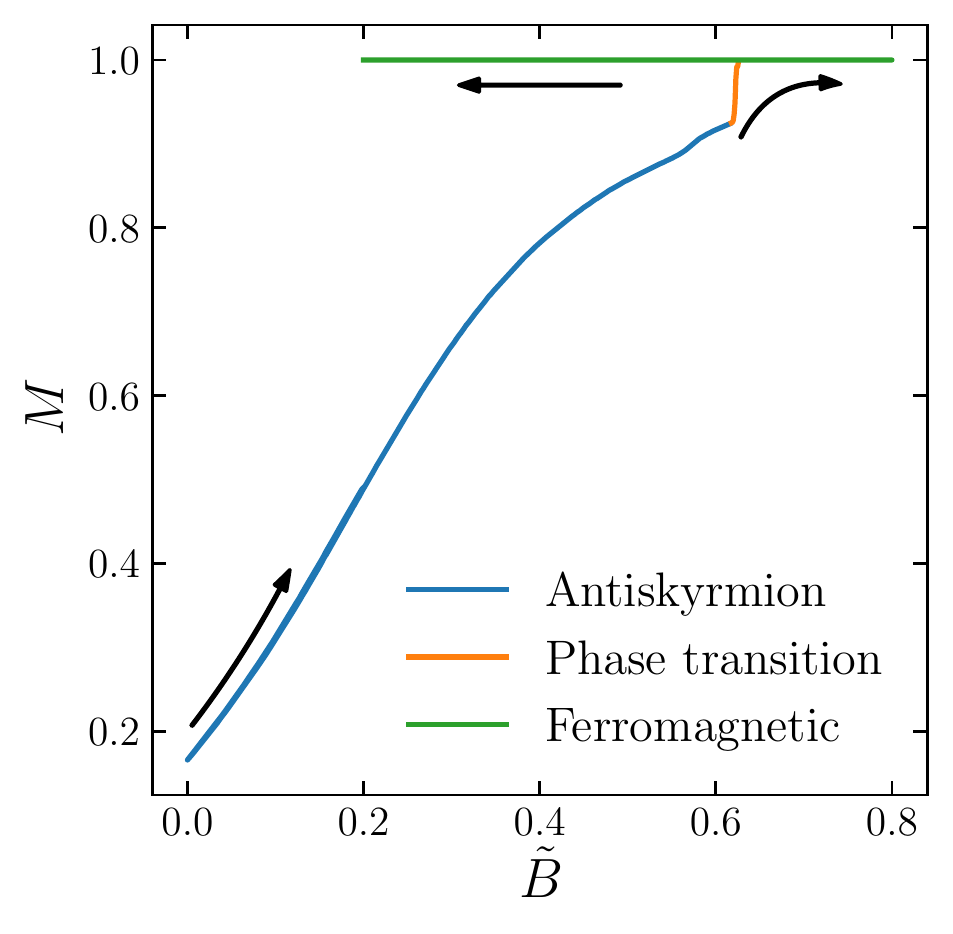}
  \hspace{1em}
  \includegraphics[width=0.635\columnwidth]{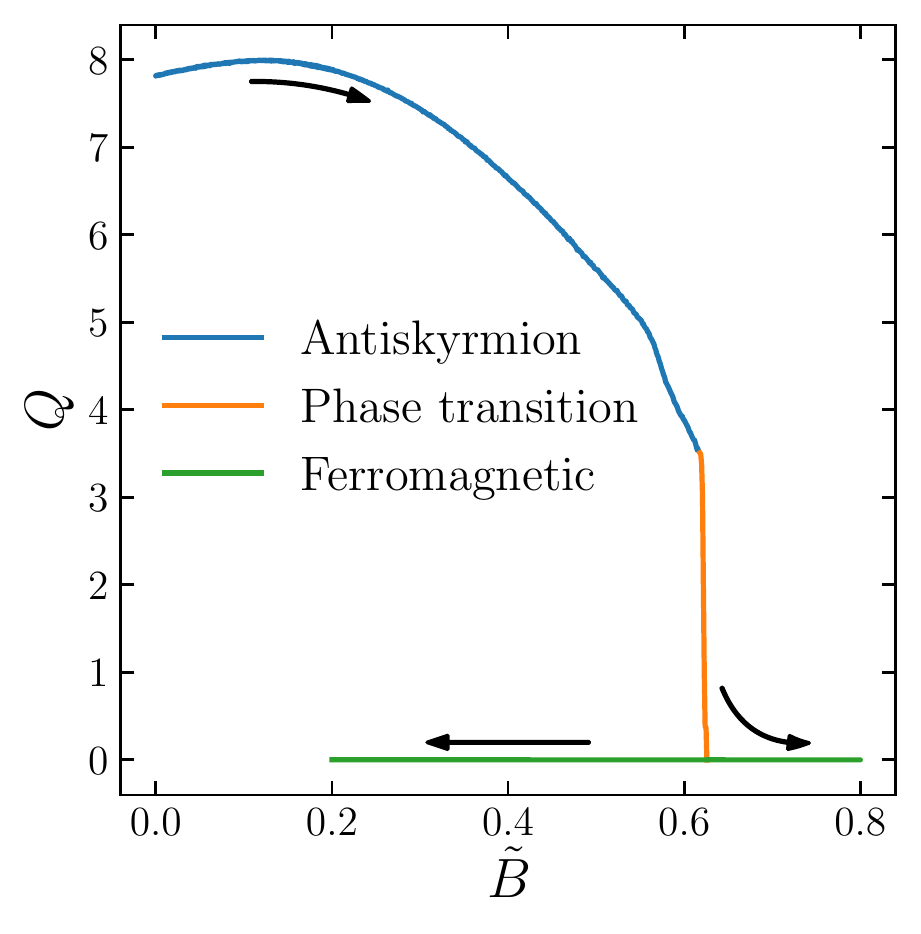}
  \hspace{1em}
  \includegraphics[width=0.65\columnwidth]{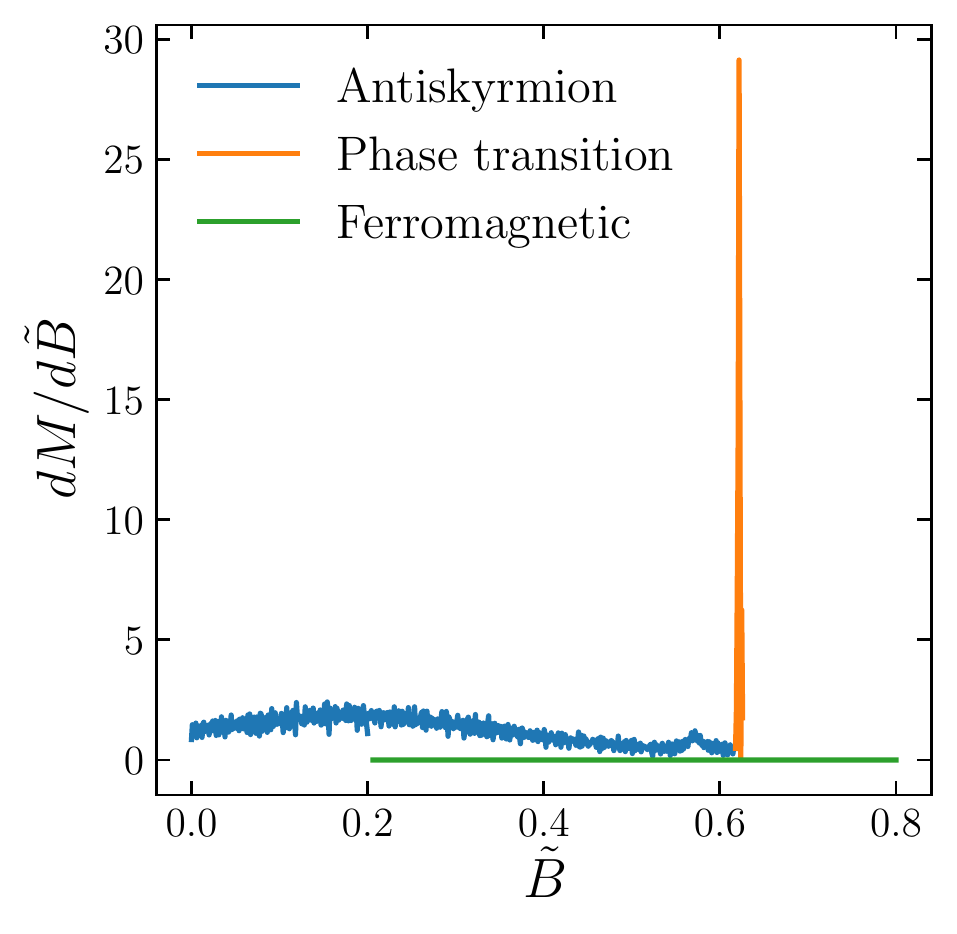}
  \caption{Magnetization $M$, topological charge $Q$ and derivative of the magnetization $dM/d\tilde{B}$ as a function of $\tilde{B}$, for the field loop schedule, defined in Section~\ref{sec:gibbs}, at $\tilde{T} = 0.001$ and for $D_{2d}$-type DM interactions.
  The small peak in $dM/d\tilde{B}$ and the quick change in $Q$ at around $\tilde{B} \simeq 0.55$ is an early sign of the phase transition, corresponding to a local collapse of some of the anti-skyrmion structures, which leads to a small region in the lattice being in the ferromagnetic state, in co-existence with the larger anti-skyrmion phase.}
  \label{fig:Hysteresis}
\end{figure*}

The presence of metastable states is a consequence of the non-convex nature of the Hamiltonian, which has several local minima separated by barriers that prevent the system from transitioning from one minimum to another at zero temperature.
The height of these barriers, therefore, determines the metastability (and hence the lifetime) of the different states in the low-temperature regime.
The same situation arises at higher temperatures, with the role of the energy being played by the Gibbs free energy, which is defined as the Legendre transform of the Helmholtz free energy,\footnote{In a quantum field theory scenario, one can think of the Helmholtz free energy as the generating functional for connected Green's functions while the Gibbs free energy corresponds to the quantum-effective action.}
\begin{equation}
  G \left[\Sb\right] = F \left[\vec{J}(\Sb)\right] + \Sb \cdot \vec{J}(\Sb) \, .
\end{equation}
Here, $\vec{J}(\Sb)$ is the solution to the equation
\begin{equation}
  \ev{\vec{S}}_{\vec{J}(\Sb)} = \Sb \, ,
\end{equation}
and the argument $\Sb$ behaves as a background field, with thermal effects causing the spins to fluctuate around it.
Clearly, for a vanishing source, $\vec{J} = 0$, the Helmholtz and Gibbs free energies are equal.
Indeed, the Gibbs free energy has the following properties that make it effectively a version of the Hamiltonian $H_d$ \emph{including} thermal corrections.
\begin{enumerate}
  \item As $T \to 0$, the path integrals reduce to their saddle-point approximation, and therefore $G\left[\Sb\right] \to H\left[\Sb\right]$.
  \item $G$ has local minima at configurations of $\Sb$ corresponding to the different (meta)stable macrostates with vanishing source, since $\md G/\md\Sb = 0$ implies $\vec{J} = 0$.
  \item The Eyring equation~\eqref{eq:eyring} describes the relation between the lifetime of metastable states and Gibbs free energy barriers.
\end{enumerate}

In this section, we aim to compute the profile of $G$ along a path in the background field space of $\Sb$ connecting the topological phase and the $xy$-translation invariant phase, i.e.~the conical state for $T/O$ or the ferromagnetic one for $D_{2d}$.
This will allow us to estimate the height of the energy barrier, $G_a$, between the two and, therefore, in turn, to determine the lifetime of the topological phase.

To construct the desired path in the integration, we turn on a constant source term,
\begin{equation}
  \vec{J} = -J_0 \unitvec{z} \, .
  \label{eq:constant-source}
\end{equation}
Using this and starting at the topological phase, a slow increase of the source from $J_0=0$ to a high value ($J_0 \simeq 0.7$, for example), followed by a decrease back to $J_0=0$, induces a phase transition into the conical phase for $T/O$ crystals and into the ferromagnetic phase for $D_{2d}$, respectively.
Thus, in both cases, one can obtain a path connecting the topological phase and the $xy$-translation invariant phase with the same procedure.

We remark that, in this setting, the thermal averages of operators and the Helmholtz free energy can be viewed as functions of temperature, magnetic field and source, denoted by $\ev{\mathcal{O}}_{\tilde{T}, \tilde{B}, J_0}$ and $F (\tilde{T}, \tilde{B}, J_0)$, respectively.
A convenient property of Eq.~\eqref{eq:constant-source} is that the source it defines can be absorbed in the magnetic field, i.e.
\begin{align}
  \ev{\mathcal{O}}_{\tilde{T}, \tilde{B}, J_0}
  &= \ev{\mathcal{O}}_{\tilde{T}, \tilde{B} + J_0,0} \, ,
  \label{eq:absorb-ev}\\
  F\left(\tilde{T}, \tilde{B}\right)\left[J_0\right] &= F \left(\tilde{T}, \tilde{B} + J_0\right)[0] \, .
\end{align}
These relations allow us to simulate both objects using the algorithm described in Section~\ref{sec:simulations}.
In view of them, the source $J_0$ can be regarded as an additional magnetic field $B$. However, we keep them separated in order perform the Legendre transform over $J_0$ for the calculation of $G$ at any given value of $B$.

The Gibbs free energy depends on the background field only through the average local magnetisation,
\begin{equation}
  M\left(\tilde{T}, \tilde{B}\right) =
  \ev{\left(\vec{S}_{\vec{r}}\right)_z}_{\tilde{T}, \tilde{B}, J_0} \, .
\end{equation}
Therefore, it can be viewed as a function of these three parameters too, $G(\tilde{T}, \tilde{B})[M]$.
Note that the source is also a function of temperature and magnetic field, which will become clear momentarily.
The Gibbs free energy has the property that $\partial G/\partial M = J_0$, which can be integrated to generate the profile of $G$ as a function of $M$ at constant temperature and magnetic field, $\tilde{T}$ and $\tilde{B}$,
\begin{equation}
  G\left(\tilde{T}, \tilde{B}\right)[M] = G_0 + \int_{M_0}^M J_0 \, \md M \, .
  \label{eq:thermal-integration}
\end{equation}
This means that $G$ can be reconstructed, up to a constant shift, from the source $J_0$.
To determine the latter in practice, one can slowly vary $J_0$, and compute the magnetisation $M$ for each source value.
That is, repeating this for every desired value of $\tilde{T}$ and $\tilde{B}$, a magnetisation function $M(\tilde{T}, \tilde{B})[J_0]$ is obtained.
The resulting relation between $J_0$ and $M$ can then be inverted.
Indeed, by means of Eq.~\eqref{eq:absorb-ev}, one only needs to compute $M(\tilde{T}, \tilde{B} + J_0)[0]$ to obtain $M(\tilde{T}, \tilde{B})[J_0]$.

To compute the magnetisation for the relevant range of parameter values, we perform the following simulations, recording $M$ along the way:
\begin{enumerate}
  \item Initialise the system on the topological phase at some initial temperature and magnetic field, $\tilde{T_0}$ and $\tilde{B}_0$, using the FC schedule.
  \item Decrease $\tilde{B}$ to 0 in 100 steps $\Delta \tilde{B}$ (here we choose $\Delta \tilde{B}_i = 0.02$ for each step $i$).
  \item Increase $\tilde{B}$ in 800 steps until a sufficiently high field ($\tilde{B}=0.8$) is reached, where the configuration has collapsed to the ferromagnetic phase.
  \item Decrease the field back to $\tilde{B}_0$ in 100 steps, again recording the observables.
\end{enumerate}
We refer to the schedule here as a \emph{field loop}.
We perform this procedure for $\tilde{B}_0 = 0.2$ and $\tilde{T}_0$ between 0.001 and 0.6, in steps of 0.05.

In the above algorithm, the integral~\eqref{eq:thermal-integration} can then be approximated using finite differences,
\begin{equation}
  G\left(\tilde{T}_0, \tilde{B}\right)_n = G\left(\tilde{T}_0, \tilde{B}\right)_0 + \sum_{i=1}^n \left(\tilde{B}_i - \tilde{B}\right) \left(M_i - M_{i - 1}\right) \, ,
  \label{eq:numerical-thermal-integration}
\end{equation}
where $\tilde{B}_i$ and $M_i$ are the values of $\tilde{B}$ and $M$ at the $i$-th step of the field loop.
Carefully note that this equation may be used only to compute the Gibbs free energy for a series of consecutive points belonging to the same macrostate.
The reason for this is illustrated in Fig.~\ref{fig:Hysteresis}, where we display $M$, $Q$ and $dM/d\tilde{B}$ as functions of $\tilde{B}$ in an example scenario.
We find that the numerical derivative of the magnetisation suddenly diverges when the system evolves to a different phase along the loop schedule, indicating a first-order phase transition.
Since the phase transition arises in jumps, through the disappearance of an integer number of (anti)skyrmions in each jump, the derivative of the magnetisation varies wildly in a relatively small section of the loop until the phase transition is completed.
Eq.~\eqref{eq:numerical-thermal-integration} can only generate the relative values of $G$ for a series of consecutive points in the same phase.
Along these lines, to assemble different series corresponding to different phases, we use the fact that $\Delta G$ equals $\Delta F$ at its minima, $\Delta G = \Delta F$.
The latter can be computed by the method described in Section~\ref{sec:hierarchy}.
We thus shift all the points in each series of points by the same amount for the difference between minima to match the value computed from $\Delta F$.

\begin{figure*}
  \centering
  \includegraphics[width=0.85\textwidth]{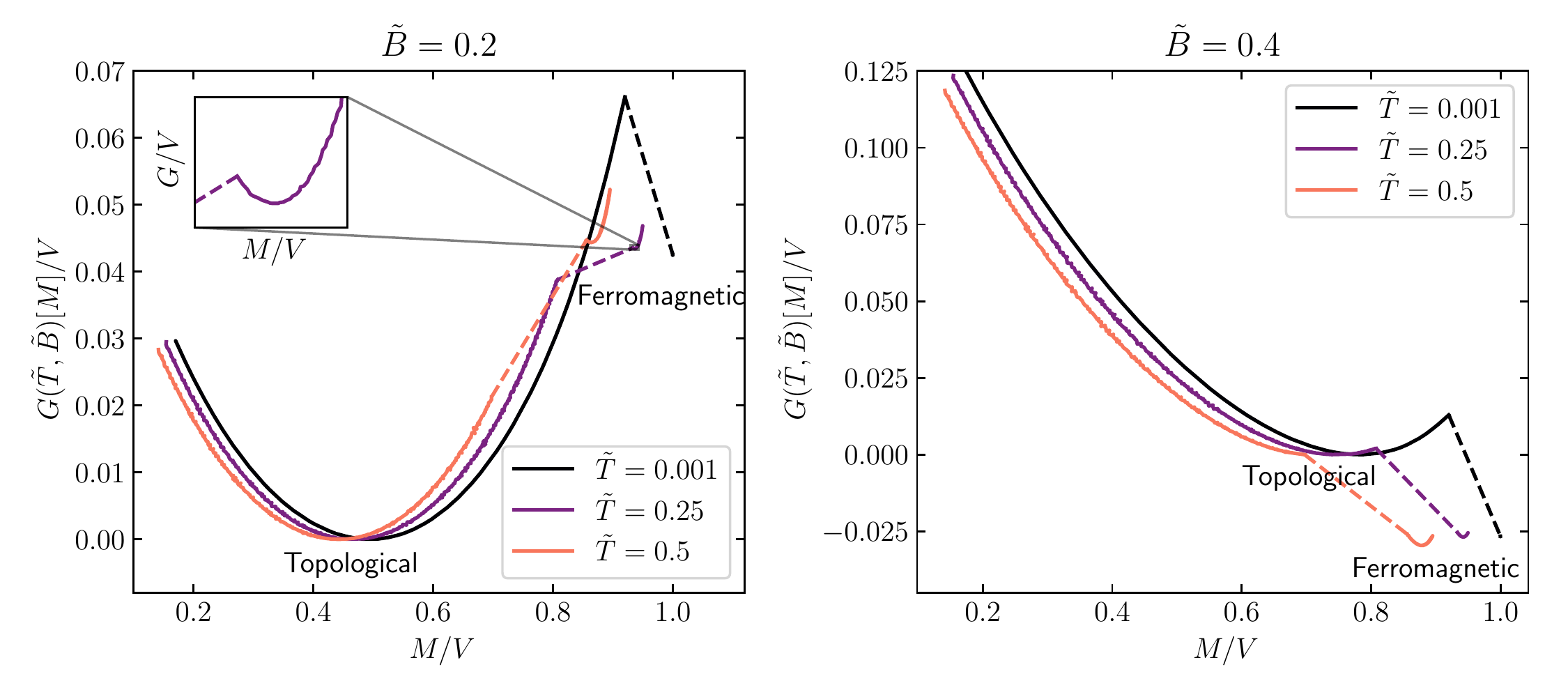}
  \caption{Example profiles of the Gibbs free energy density $G / V$ as a function of the magnetisation density $M / V$ for different temperatures and $\tilde{B} = 0.2$ (left) and $\tilde{B} = 0.4$ (right), with DM interactions of $D_{2d}$ type.
    These profiles have been generated through numerical integration along the loop schedule described in Section~\ref{sec:gibbs}.
    The local minima correspond to the topological and ferromagnetic phases.
    The inset axes on the left zooms in to show the ferromagnetic local minimum.
    The dotted line represents the piece of the path in which $M$ has discontinuities, signalling a first-order phase transition between these two phases.}
  \label{fig:GvsM}
\end{figure*}

\begin{figure*}
	\centering
	\includegraphics[width=0.65\columnwidth]{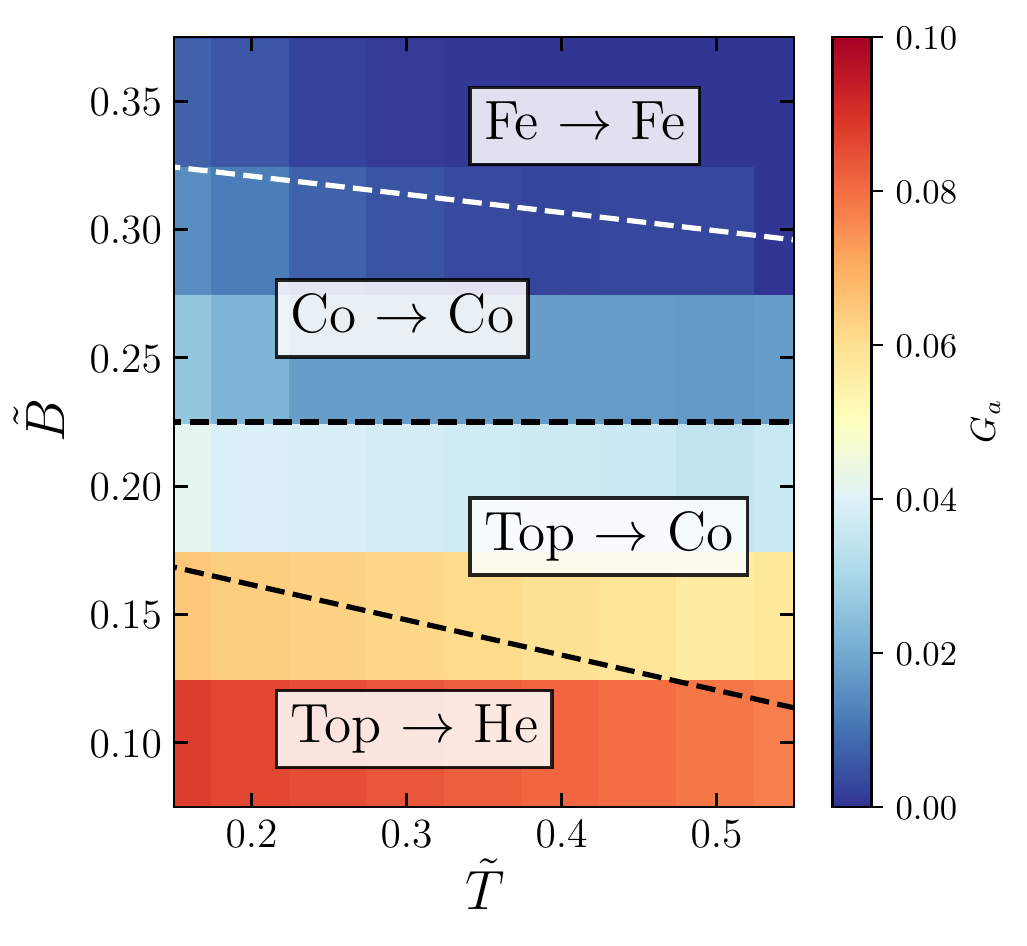}
	\hspace{1em}
	\includegraphics[width=0.65\columnwidth]{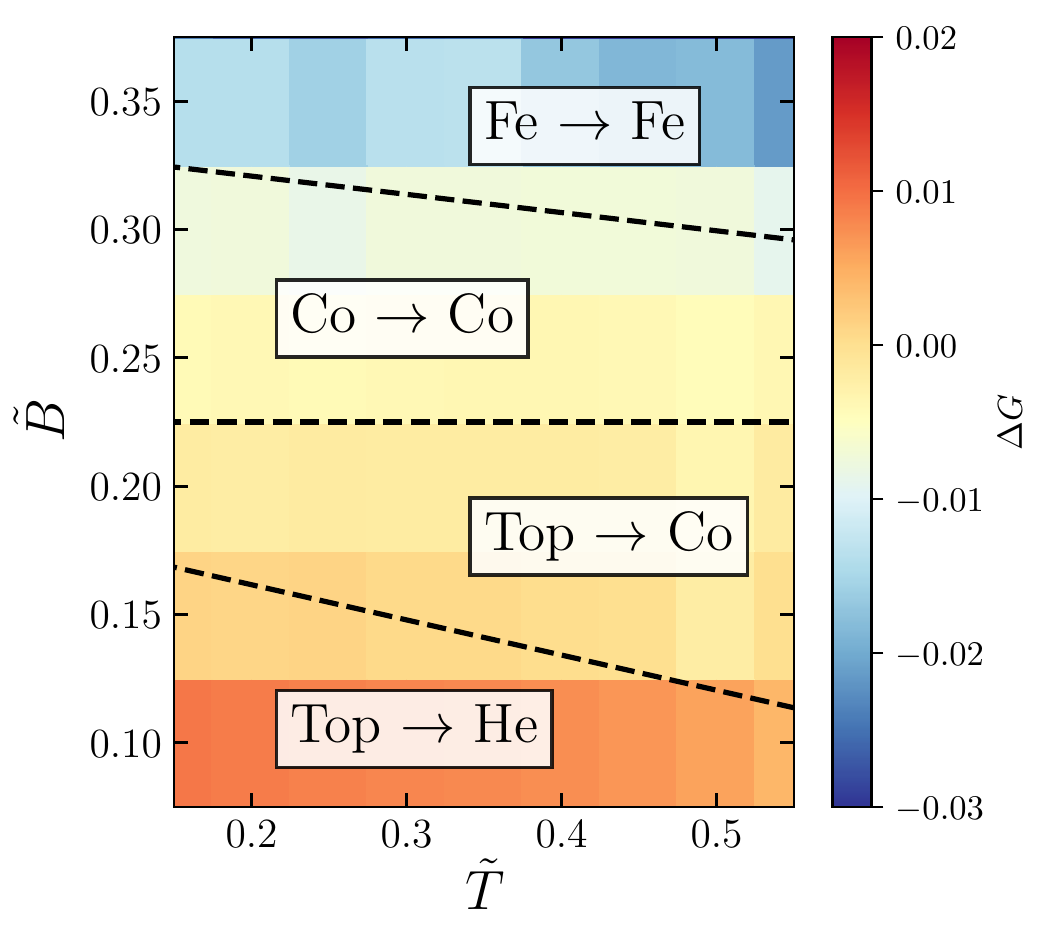}
	\hspace{1em}
	\includegraphics[width=0.65\columnwidth]{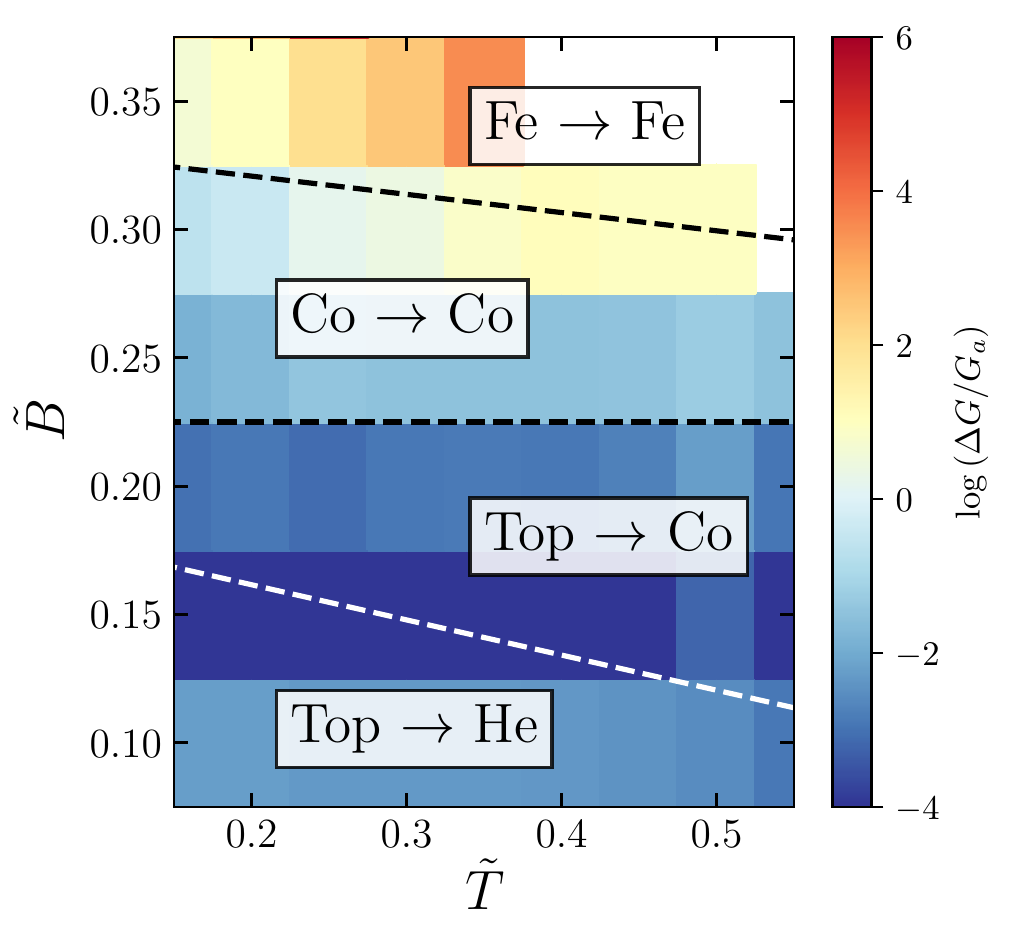}
	\includegraphics[width=0.65\columnwidth]{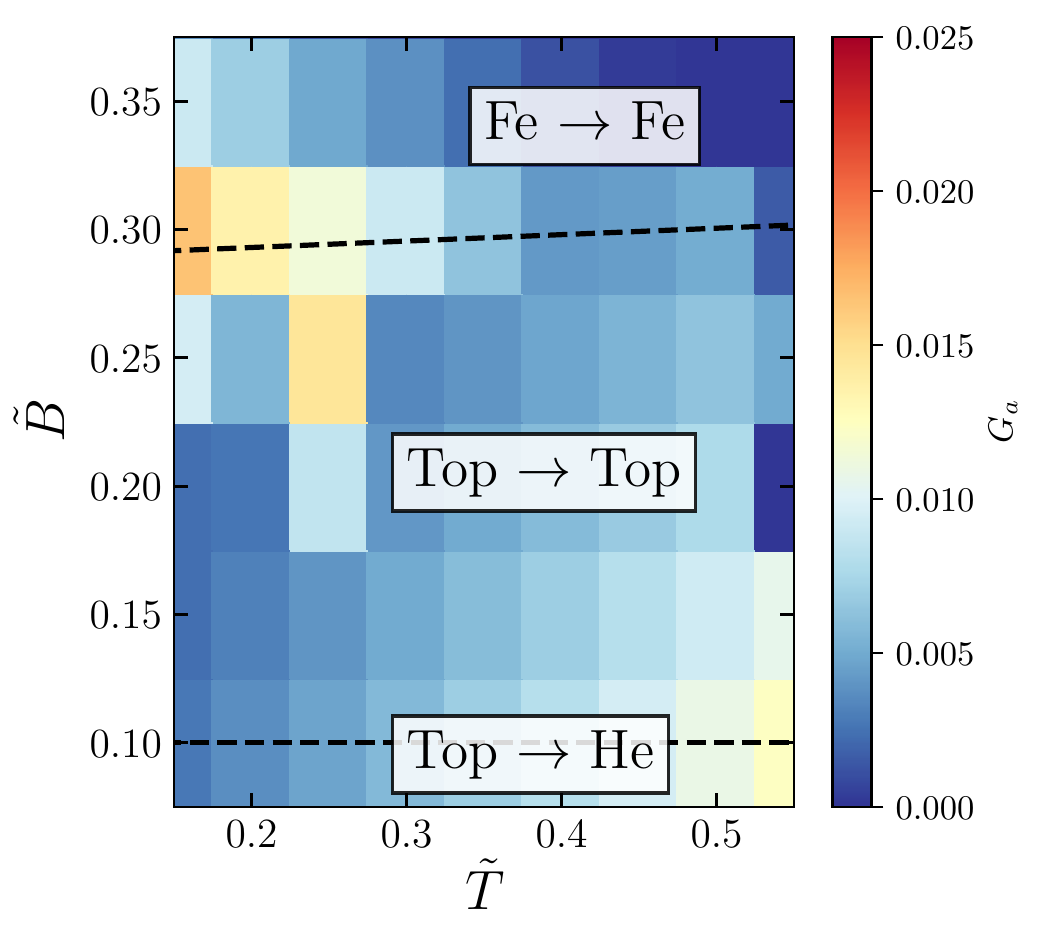}
	\hspace{1em}
	\includegraphics[width=0.65\columnwidth]{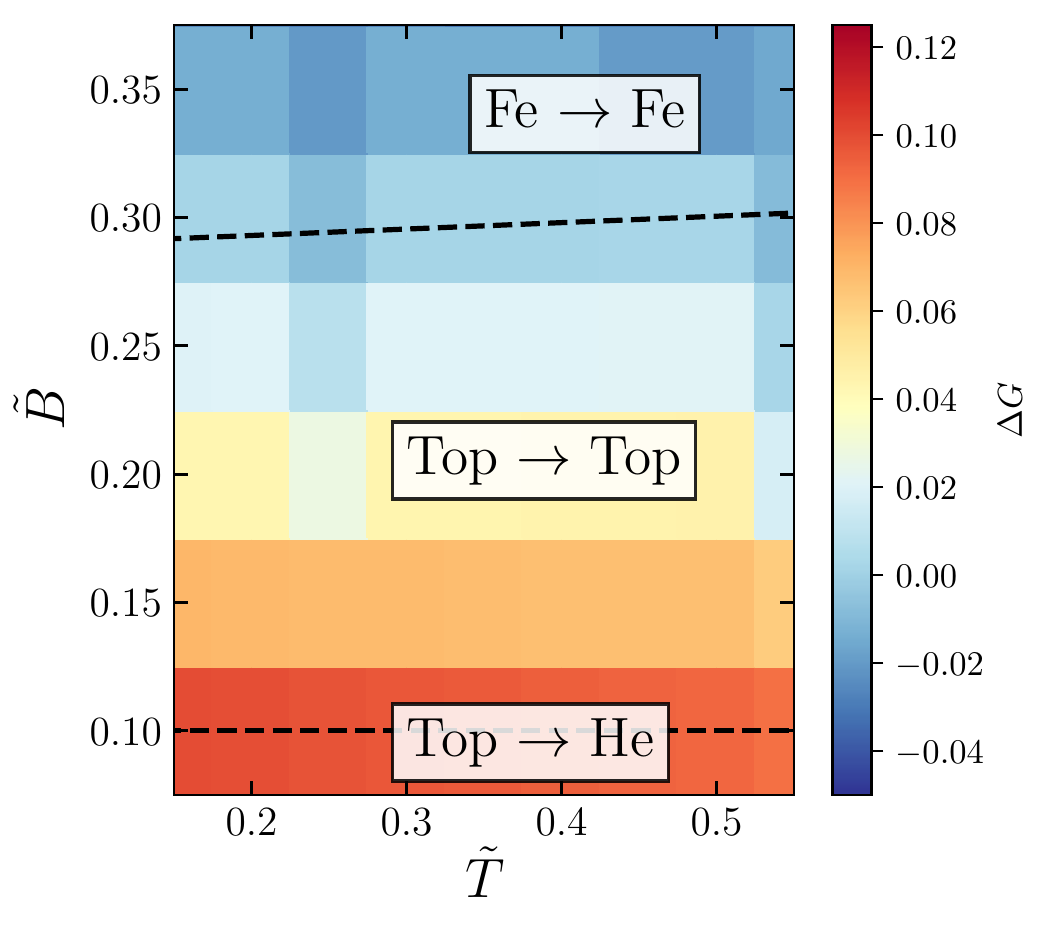}
	\hspace{1em}
	\includegraphics[width=0.65\columnwidth]{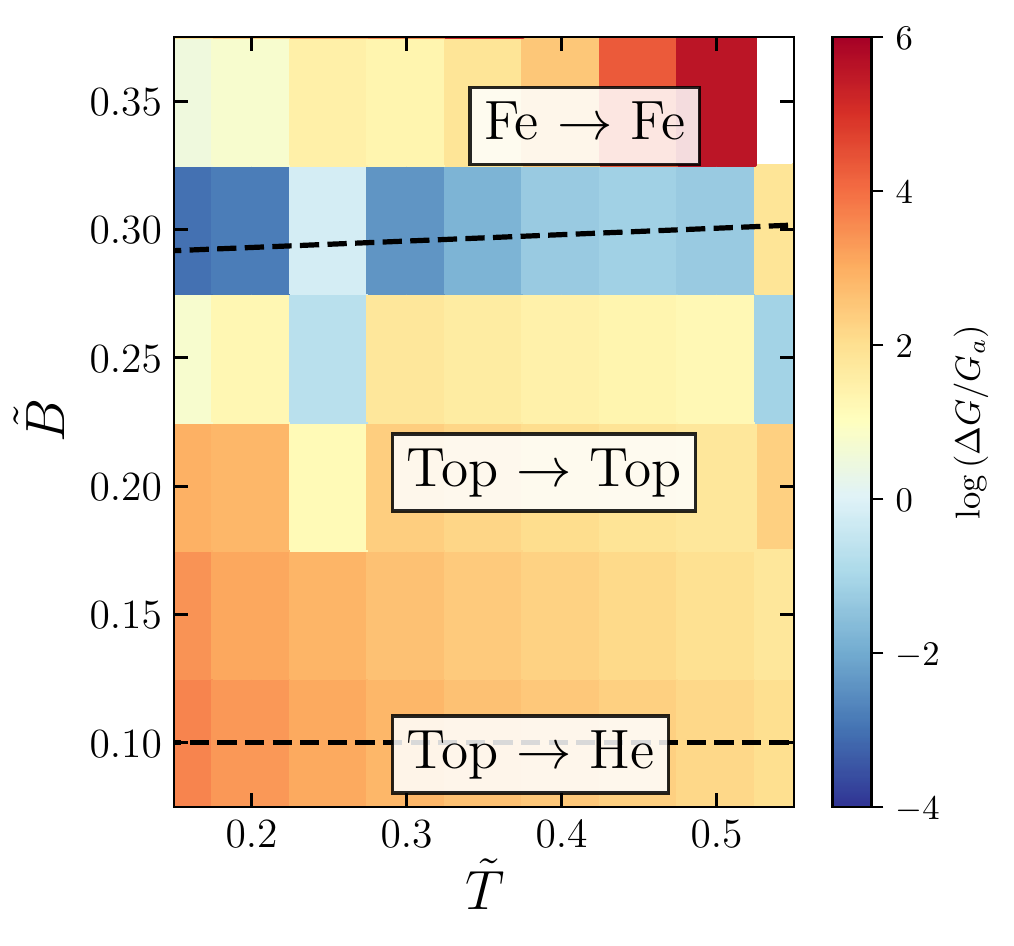}
	\caption{Activation energy $G_a$ (left), free-energy difference $\Delta G$ (centre), and their ratio $\Delta G / G_a$ (right) as a function of temperature and magnetic field, $\tilde{T}$ and $\tilde{B}$. The upper row illustrates a $T$ or $O$ symmetric crystal (featuring Bloch skyrmions), while the lower one illustrates a $D_{2d}$ system (featuring antiskyrmions). These quantities are associated to the transition between metastable states and the true ground state. The states are either conical (Co), topological (Top), helical (He) or ferromagnetic (Fe), or even a nontrivial superposition of these. Rough estimates for boundaries between the different phase transitions are indicated by the dashed lines. The rightmost panel is included for experimental indication (see main text). Here, the white region shows the divergence of the ratio due to a vanishing barrier height at high temperatures and large magnetic fields.}
	\label{fig:G_a}
  \end{figure*}

In Fig.~\ref{fig:GvsM}, we show some example profiles of the Gibbs free energy along a loop schedule constructed using the above procedure.
As a cross-check, we have verified the crucial property that $G$ approaches $\ev{H}$ at low temperatures.
The antiskyrmion phase is the ground state at $\tilde{B} = 0.2$, and metastable at $\tilde{B} = 0.4$, with the ferromagnetic phase playing the opposite role.

\begin{figure*}
  \centering
  \begin{minipage}{0.25\linewidth}
    \includegraphics[width=\textwidth]{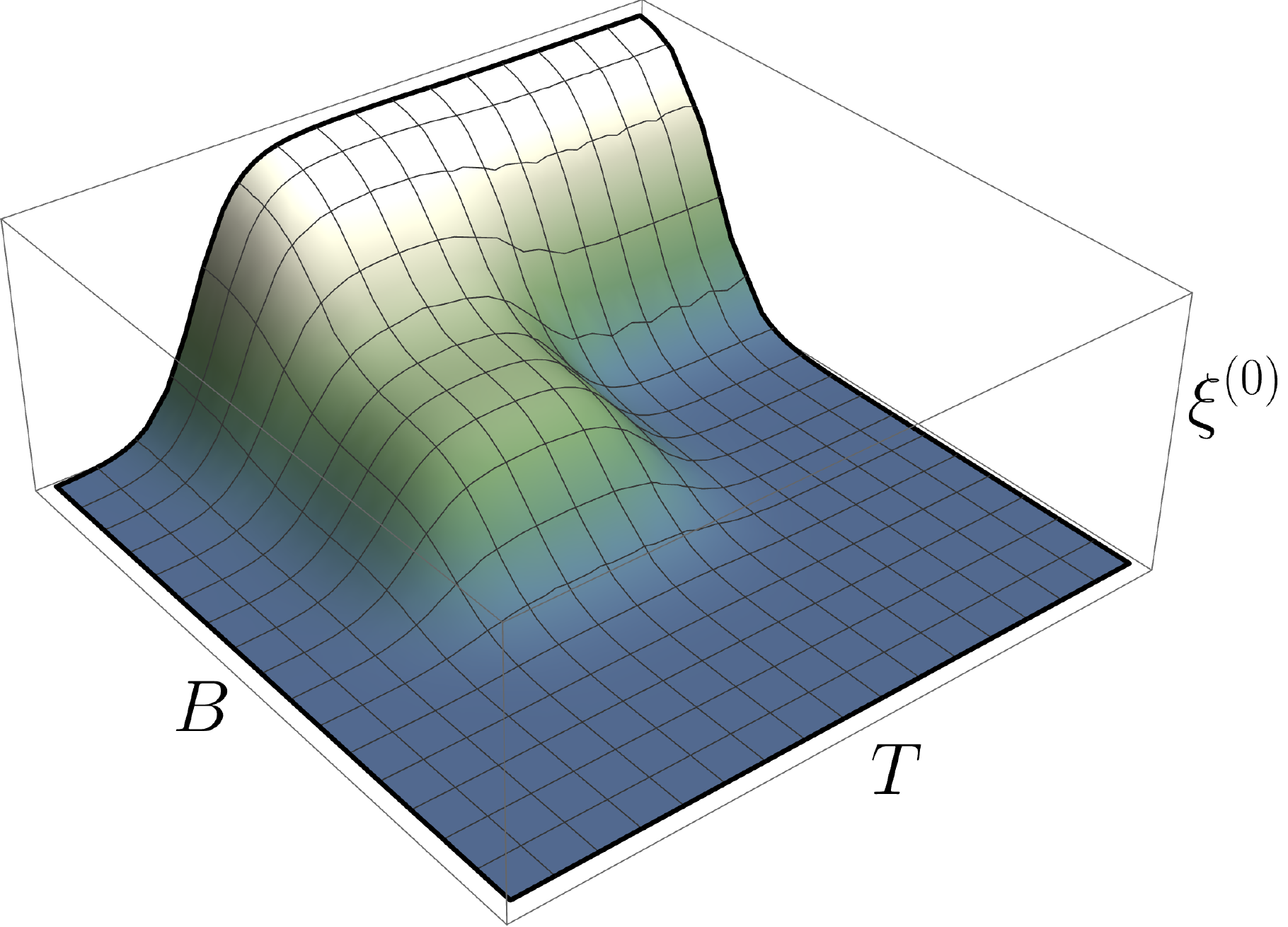}
    \includegraphics[width=\textwidth]{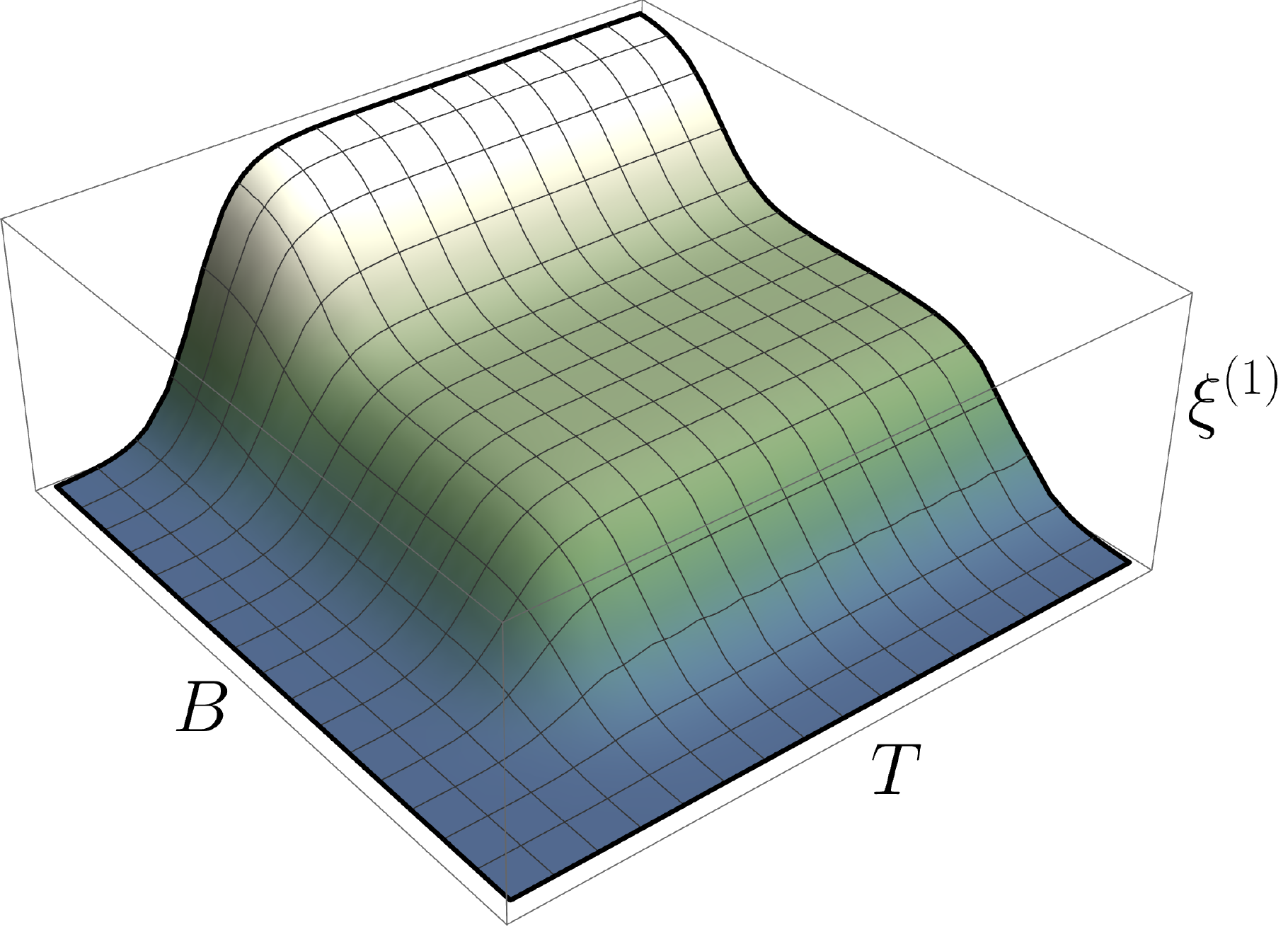}
    \includegraphics[width=\textwidth]{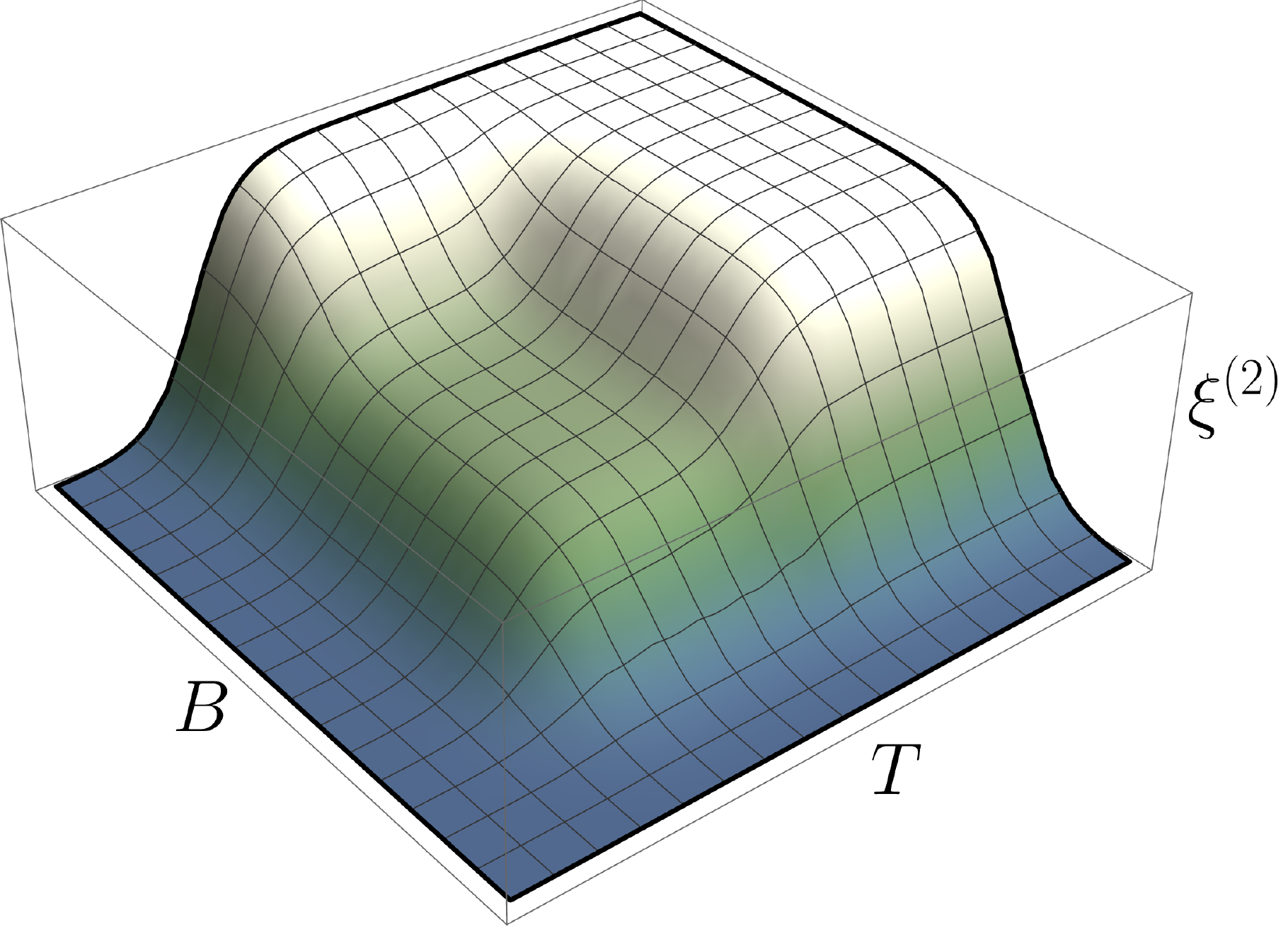}
  \end{minipage}
  \begin{minipage}{0.74\linewidth}
    \includegraphics[width=\textwidth]{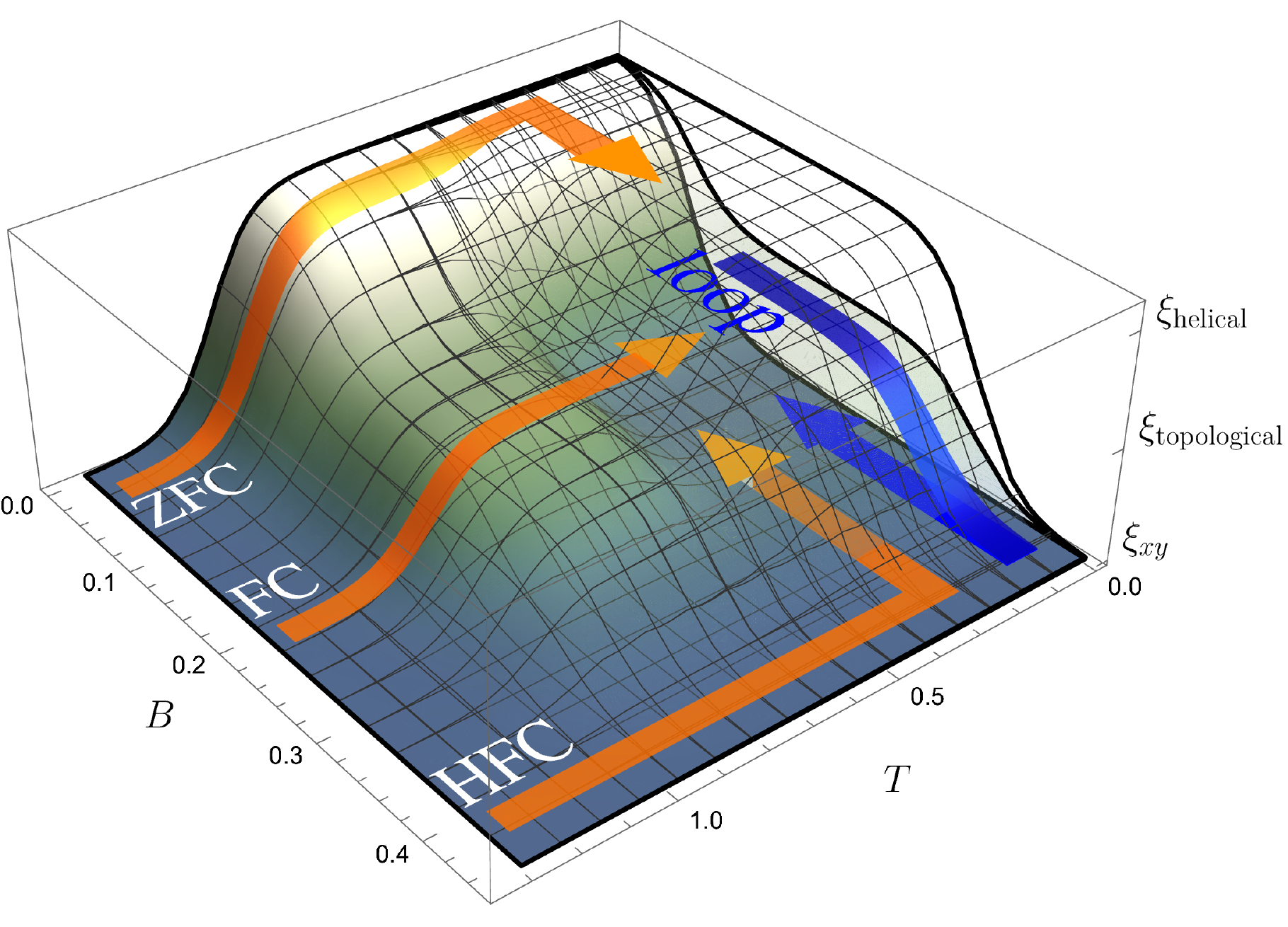}
  \end{minipage}
  \caption{Schematic representation of the 3 sheets $\xi^{(0)}(\tilde{T}, \tilde{B})$, $\xi^{(1)}(\tilde{T}, \tilde{B})$ and $\xi^{(2)}(\tilde{T}, \tilde{B})$ of the surface of free minima of the Gibbs free energy $G$. Left: individual representation for all sheets. Right: All sheets were drawn together, with arrows indicating the paths in $(\tilde{T}, \tilde{B}, \xi)$ space through the different schedules we consider in this work.}
  \label{fig:sheets}
\end{figure*}

In this scenario, we can estimate the value of the energy barrier between phases, $G_a$, by taking the maximum value of $G$ along the piece of the loop between the two local minima.
While Fig.~\ref{fig:GvsM} illustrates these energy barriers for specific values of the external field, $\tilde{B}=0.2$ and $\tilde{B}=0.4$, respectively, our field-loop approach is able to chart the Gibbs free energy landscape over the entire phase diagram in temperature and magnetic field.
We illustrate this in Fig.~\ref{fig:G_a}.
Here, we perform the field loop mentioned before for (metastable) states at different temperatures and external field.
These states are prepared through an FC schedule.
As an example, we study the low-temperature regime, $0.15 \lesssim \tilde{T} \lesssim 0.55$, where, depending on the magnetic field, the metastable states are conical, topological, or helical or a superposition of these (see third column Fig.~\ref{fig:phase-diagram}).
Performing the field loop will then drive the system to the true ground state, shown in the rightmost panels of Fig.~\ref{fig:phase-diagram}.
In Fig.~\ref{fig:G_a}, we track this transition and illustrate the energy barrier between these phases, $G_a$, together with the energy difference between vacua, $\Delta G = \Delta F$, computed via the method described in Section~\ref{sec:hierarchy}.
Rough estimates for boundaries between the different phase transitions are indicated by the dashed lines.
We also show the ratio $\Delta G / G_a$ as a quick indication of what could be measured in an experiment, which we will briefly comment on later.

Clearly, for both crystal structures at high magnetic fields and large temperatures, the ferromagnetic phase dominates and the energy barrier between vacua vanishes.
Similarly, for a $T$ or $O$ symmetric structure, featuring Bloch skyrmions, the free-energy difference grows for small magnetic field, indicating that the dominant helical phase in this region is well separated from other vacua, further reinforced by the right panel of the figure.
This showcases the potential of our method to determine the separation between vacua everywhere in the thermodynamical phase diagram.

We note that our results for the activation energy and free-energy difference illustrated in Fig.~\ref{fig:G_a} are given as dimensionless quantities, that do not have a physical meaning \emph{a priori}.
At the same time, in principle, the results we have obtained in this section can be tested experimentally in materials in which the ferromagnetic and DM interactions dominate.
A potential procedure for this would be to prepare the system in the topological phase in the region where it is metastable through an FC schedule and then perform a loop schedule, tracking the magnetisation along it.
The activation energy $G_a$ and free-energy difference $\Delta G$ can be computed in physical units using Eq.~\eqref{eq:numerical-thermal-integration}.
The ratio $\Delta G / G_a$ should agree with the one for the dimensionless version of both quantities we have shown in this work.

\section{Predicting the evolution for arbitrary schedules}
\label{sec:predicting}

In the previous sections, we have established that several phases can exist at the same temperature and magnetic field, $\tilde{T}$ and $\tilde{B}$, corresponding to the different minima of the Gibbs free energy, $G(\tilde{T}, \tilde{B})$.
We now want to address the question of how to predict which of these phases is chosen by the system at the end of some process, using as the input data the initial state of the system and the path it follows in $(\tilde{T}, \tilde{B})$ space.

We assume the existence of a continuous function $\xi(\Sb)$ in the space of $\Sb$ configurations that takes different values $\xi_{xy} < \xi_{\text{topological}} < \xi_{\text{helical}}$ at the different phases, where $\xi_{xy}$ corresponds to the $xy$-translation invariant phase (conical, for $T/O$ crystals, or ferromagnetic, for $D_{2d}$).
The set of minima of $G(\tilde{T}, \tilde{B})[\Sb]$ form a surface in $(\tilde{T}, \tilde{B}, \xi)$ space with three sheets,
\begin{equation}
  \xi^{(0)}(\tilde{T}, \tilde{B}) \leq \xi^{(1)}(\tilde{T}, \tilde{B}) \leq \xi^{(2)}(\tilde{T}, \tilde{B}).
\end{equation}
At a $(\tilde{T}, \tilde{B})$ space in which there is only one possible phase, we have $\xi^{(0)} = \xi^{(1)} = \xi^{(2)}$.
A schematic representation of this surface is shown in Fig.~\ref{fig:sheets}.

Throughout the 4 schedules we have considered so far (ZFC, FC, HFC and loop), the evolution of the system consistently shows a tendency to stay in the same $\xi^{(i)}$ sheet unless that sheet ceases to exist for the given $(\tilde{T}, \tilde{B})$. More concretely, the path that the system follows in $(\tilde{T}, \tilde{B}, \xi)$ space can be predicted from the initial state by following two rules:
\begin{enumerate}
  \item $\xi$ is continuous as one moves along the path.
  \item When $\xi$ branches at a point, the evolution follows along the current sheet.
\end{enumerate}
The states predicted by these rules match those we find in our simulations in all four schedules in both types of crystals. In order to test them in a different scenario, we consider a new schedule, defined by performing an HFC to the region where a conical (for $T$ or $O$ crystal) or ferromagnetic (for $D_{2d}$) state is reached, with the final $\tilde{B}$ in the region where the topological phase is the ground state at higher temperatures, and then increasing the temperature.
Following the rules, the system will reach the topological phase, which agrees with what we find in the previous simulations.
We have thus shown that they work across all the boundaries of the regions in $(\tilde{T}, \tilde{B})$ space corresponding to the different phases, for multiple points of crossing and various previous histories of the system.

Finally, we conjecture that these rules, which we have derived for the quasi-static evolution of the system under Monte Carlo simulations, will hold for materials dominated by ferromagnetic and DM interactions.
This is supported by the fact that such simulations have been shown to correctly reproduce the phase diagram of Bloch skyrmions~\cite{Buhrandt:2013uma}.
Again, these rules can be tested experimentally by preparing the system on each available state at some external parameter point, changing the temperature or magnetic field to make it go through the branching regions, and observing the phase transitions.

\section{Conclusions}
\label{sec:conclusions}

In this work, we have studied the structure and properties of metastable states in chiral magnets using Monte Carlo simulations of a spin-lattice system with ferromagnetic and DM interactions.
Depending on the path followed by the system, as the temperature and external magnetic field are varied, the final state can be in different thermodynamical phases for the same initial state and final parameter values of $\tilde{T}$ and $\tilde{B}$.
In fact, this leads to a multi-valued free-energy function, with one branch per metastable state.

We have implemented a general method for computing free-energy differences between metastable states of the system.
By this method, we have found the ordering in the free energies of the different states, and, using these results, we have obtained a complete phase diagram providing the ground state at every value of the temperature and external magnetic field for both types of DM interactions we consider.
Furthermore, we have determined the free-energy difference $\Delta F$ between the topological phase and the $xy$-translation invariant phase, which is conical or ferromagnetic, depending on the DM interaction type.

Crucially, we have also proposed a procedure for computing the Gibbs free energy, which plays the role of the Hamiltonian function, including thermal corrections, along a path connecting the topological phase and the $xy$-translation invariant phase.
In practice, we have successfully employed this procedure to obtain the energy barrier, i.e.~the, activation energy $G_a$, separating both phases (at a distance $\Delta G$).
In principle, these quantities can also be measured experimentally in a similar way.
We expect our numerical results for the ratio $\Delta G / G_a$ to provide a robust prediction for the values obtained through the experiment.
Determining $G_a$ is relevant in technological applications because it controls the lifetime of the corresponding (anti)skyrmion structures when metastable.

Finally, we have provided a set of rules to predict the thermodynamical phases a chiral magnet will go through when following an arbitrary schedule.
According to these rules, one can use the map we have generated for the three-sheeted surface of the Gibbs free energy's local minima and assume continuity and the system's tendency to stay on the same sheet to get a well-defined prediction for every schedule.
Similar to the activation energy, we expect that these rules can be robustly tested in an experiment.
They can potentially be used to prepare a chiral magnet in any available state or even trigger any desired phase transition between thermodynamical phases.

\phantom{paragraph}

\acknowledgments

This work was supported by the UK Skyrmion Project EPSRC Programme Grant (EP/N032128/1).
J.~C.~C.~is supported by the Spanish Ministry of Science and Innovation, under the Ramón y Cajal program.
S.~S.~is funded by the Deutsche Forschungsgemeinschaft (DFG, German Research Foundation) -- 444759442.

\appendix

\section{Next-to-nearest neighbor corrections}
\label{sec:nnn-corrections}

The microscopic model $H_d$ (Eq.~\eqref{eq:discrete-hamiltonian}) approximates the coarse-grained system described by $H$ (Eq.~\eqref{eq:hamiltonian}) for small lattice spacing $a$, with a relative error of order $a^2$.
This can be improved to a relative error of order $a^4$ by introducing next-to-nearest neighbors interactions with the same structure, as shown in Ref.~\cite{Buhrandt:2013uma}.
We provide here an alternative proof based on the Taylor expansion
\begin{equation}
  M_s (\vec{S}_{\vec{r} + a \vec{e}_i} - \vec{S}_{\vec{r}})
  =
  \sum_{n = 1}^\infty \frac{a^n}{n!} \, \partial_i^n\vec{M}
\end{equation}
where $\vec{e}_1 = \vec{x}$, $\vec{e}_2 = \vec{y}$ and $\vec{e}_3 = \vec{z}$.

The lattice approximation to the ferromagnetic exchange term can be derived from
\begin{align}
  M^2_s \sum_i (\vec{S}_{\vec{r} + a \vec{e}_i} - \vec{S}_{\vec{r}})^2
  &=
  a^2 (\nabla \vec{M})^2
  + \frac{a^4}{4} \sum_i (\partial_i^2 M)^2
  \nonumber \\
  &\phantom{=}
  + [\text{odd}]
  + O(a^6),
  \label{eq:finite-difference}
\end{align}
where $[\text{odd}]$ denotes terms with an odd number of derivatives, which vanish when integrated over $\vec{r}$ with periodic boundary conditions.
Neglecting these terms, we have
\begin{equation}
  \frac{M^2_s}{a^2} \sum_i (\vec{S}_{\vec{r} + a \vec{e}_i} - \vec{S}_{\vec{r}})^2
  =
  (\nabla \vec{M})^2 + O(a^2).
\end{equation}
Expanding the left-hand side one obtains the nearest neighbors Hamiltonian $H_d$ up to constant terms.
Including next-to-nearest neighbors interactions with the same structure a suitable coefficient allows us improve this to
\begin{multline}
  \frac{4 M^2_s}{3 a^2} \sum_i \left[
    (\vec{S}_{\vec{r} + a \vec{e}_i} - \vec{S}_{\vec{r}})^2
    - \frac{1}{16} (\vec{S}_{\vec{r} + 2a \vec{e}_i} - \vec{S}_{\vec{r}})^2
  \right]
  \\
  =
  (\nabla \vec{M})^2 + O(a^4).
\end{multline}

A similar procedure can be applied to the DM interaction terms.
They can always be written as $\sum_{ijk} \mathcal{K}_{ijk} \vec{M}_i \partial_j \vec{M}_k$,
with $\mathcal{K}_{ijk} = -\mathcal{K}_{kji}$.
One then has
\begin{multline}
  M_s^2 \sum_{ijk} \mathcal{K}_{ijk} (\vec{S}_{\vec{r}})_i (\vec{S}_{\vec{r} + a \vec{e}_j} - \vec{S}_{\vec{r}})_k
  \\
  =
  \sum_{ijk} \mathcal{K}_{ijk} \left(
    a \vec{M}_i \partial_j \vec{M}_k
    + \frac{a^3}{6} \vec{M}_i \partial_k^3 \vec{M}_k
  \right)
  \\
  + [\text{even}]
  + O(a^5).
\end{multline}
In this case, because of the anti-symmetry of $\mathcal{K}$, the terms $[\text{even}]$ with an even number of derivatives vanish under the $\vec{r}$ integral.
Neglecting them, one gets the following approximation with nearest neighbor interactions only
\begin{multline}
  \frac{M_s^2}{a} \sum_{ijk} \mathcal{K}_{ijk} (\vec{S}_{\vec{r}})_i
  (\vec{S}_{\vec{r} + a \vec{e}_j} - \vec{S}_{\vec{r}})_k
  \\
  = \sum_{ijk} \mathcal{K}_{ijk} \vec{M}_i \partial_j \vec{M}_k + O(a^2).
\end{multline}
The inclusion of next-to-nearest neighbor interactions of the same form with an appropriate coefficient gives:
\begin{multline}
  \hspace{-10pt}
  \frac{4M_s^2}{3a} \sum_{ijk} \mathcal{K}_{ijk} (\vec{S}_{\vec{r}})_i
  \left[
    (\vec{S}_{\vec{r} + a \vec{e}_j} - \vec{S}_{\vec{r}})_k
    - \frac{1}{8} (\vec{S}_{\vec{r} + 2 a \vec{e}_j} - \vec{S}_{\vec{r}})_k
  \right]
  \\
  = \sum_{ijk} \mathcal{K}_{ijk} \vec{M}_i \partial_j \vec{M}_k + O(a^4).
\end{multline}

The order $a^2$ corrections eliminated through this method are anisotropic in both cases.
Since the continuum Hamiltonian to be approximated is isotropic, it is particularly useful to keep the symmetry up to higher orders in the lattice version.

The practical utility of this method can be assessed by comparing the results obtained using it with the experimental observations.
It has been shown in Ref.~\cite{Buhrandt:2013uma} that the correct phase diagram for Bloch skyrmions is obtained in this way, which was not possible with just the nearest neighbors version for the typical lattice sizes we consider.
In order to ensure that there is no need to include additional finite-size effects (which would now be of order $a^4$), we have performed simulations for a larger $60 \times 60 \times 60$ lattice, at several representative points in parameter space. We have not found any noticeable differences between in them and the ones for the $30 \times 30 \times 30$ lattice that we have used in the rest of this work.

The effects of these corrections have been studied in Refs.~\cite{FALCIONI1983313,MUSTO198395}, in the context of the $O(N)$ non-linear $\sigma$-model which, for $N = 3$, coincides with the system we study at $K = B = 0$ and in 2 dimensions. It has been found that they significantly improve the convergence to the continuum limit of observables in Monte Carlo simulations. Similar ideas have been applied to other systems, such as the 3D Ising model~\cite{Hasenbusch:1998gh} and quantum chromodynamics~\cite{Luscher:1996sc}.

\bibliography{references}

\end{document}